\documentclass[preprint,review,12pt]{elsarticle}
\usepackage{epsfig}
\usepackage{amsmath}
\usepackage{amssymb}
\usepackage{amsthm}
\usepackage[boxed]{algorithm}
\usepackage{algorithmic}
\usepackage[tight,footnotesize]{subfigure}
\usepackage{calc}
\usepackage{multicol}

\usepackage[font=footnotesize,caption=false]{subfig}
\usepackage{fixltx2e}

\biboptions{comma,round,sort&compress}
\journal{Theory of Computer Science}

\newtheorem{theorem}{Theorem}
\newtheorem{result}{Result}
\newtheorem{corollary}{Corollary}
\newtheorem{lemma}{Lemma}

\theoremstyle{definition}

\theoremstyle{remark}

\numberwithin{equation}{section}

\newsavebox{\SquareEmpty}
\newcommand{\DefineSquareEmpty}  {
    \savebox{\SquareEmpty}(1,1)[l]{ %
        \multiput(0,0)(1,0){2}{\line(0,1){1}}
        \multiput(0,0)(0,1){2}{\line(1,0){1}}
    } 
} 

\newsavebox{\SquareWithX}
\newcommand{\DefineSquareWithX}  {
    \savebox{\SquareWithX}(1,1)[l]{ %
        \multiput(0,0)(1,0){2}{\line(0,1){1}}
        \multiput(0,0)(0,1){2}{\line(1,0){1}}
        \put(0,0){\line(1,1){1}}
        \put(0,1){\line(1,-1){1}}
    } 
} 

\newsavebox{\SquareFull}
\newcommand{\DefineSquareFull}  {
    \savebox{\SquareFull}(1,1)[l]{ %
        \linethickness{\unitlength} %
        \put(0.5,0){\line(0,1){1}}
    } 
} 

\newsavebox{\SquareBarsA}
\newcommand{\DefineSquareBarsA}  {
    \savebox{\SquareBarsA}(1,1)[l]{ %
        \linethickness{0.015\unitlength} %
        \multiput(0,0)(0.2,0){6}{\line(0,1){1}}
        \multiput(0,0)(0,0.2){6}{\line(1,0){1}}
    } 
} 

\newsavebox{\SquareBarsB}

\newcommand{\FourNeigh}{4 Neighbors }       
\newcommand{\EightNeigh}{8 Neighbors }      

\newcommand{\CritPoints}{\emph{critical points}}

\begin{document}

%
%
%

\begin{frontmatter}

\title{Static and Expanding Grid Coverage with Ant Robots : Complexity Results}

\author[technion,bgu]{Yaniv Altshuler\corref{cor1}}
\ead{yanival@cs.technion.ac.il}
\ead[url]{www.cs.technion.ac.il/~yanival}

\author[technion]{Alfred M. Bruckstein}
\ead{freddy@cs.technion.ac.il}
\ead[url]{www.cs.technion.ac.il/~freddy}

\cortext[cor1]{Corresponding Author (Tel : +972.544.950093 Fax : +972.153.544.950093)}
\address[technion]{Technion --- Israel Institute of Technology, Technion City, Haifa 32000, Israel}
\address[bgu]{Deutsche Telekom Laboratories, Ben Gurion University of the Negev, POB 653, Beer Sheva 84105, Israel}

\begin{abstract}
In this paper we study the strengths and limitations of collaborative teams of simple agents. In particular, we discuss the efficient use of ``ant robots'' for covering a connected region on the $\mbox{\bf Z}^{2}$ grid, whose area is unknown in advance, and which expands at a given rate, where $n$ is the initial size of the connected region.
We show that regardless of the algorithm used, and the robots' hardware and software specifications, the minimal number of robots required in order for such coverage to be possible is $\Omega({\sqrt{n}})$.
In addition, we show that when the region expands at a sufficiently slow rate, a team of $\Theta(\sqrt{n})$ robots could cover it in at most $O(n^{2} \ln n)$ time.
This completion time can even be achieved by myopic robots, with no ability to directly communicate with each other, and where each robot is equipped with a memory of size $O(1)$ bits w.r.t the size of the region (therefore, the robots cannot maintain maps of the terrain, nor plan complete paths).
Regarding the coverage of non-expanding regions in the grid, we improve the current best known result of $O(n^{2})$ by demonstrating an algorithm that guarantees such a coverage with completion time of $O(\frac{1}{k} n^{1.5} + n)$ in the worst case, and faster for shapes of perimeter length which is shorter than $O(n)$.
\end{abstract}


\end{frontmatter}

\section{Introduction}

\noindent \textbf{Motivation}.
In nature, ants, bees or birds often cooperate to achieve common goals and exhibit amazing feats of swarming behavior and collaborative problem solving. It seems that
these animals are ``programmed'' to interact locally in such a way that the
desired global behavior will emerge even if some individuals of the
colony die or fail to carry out their task for some other reasons. It is
suggested to consider a similar approach to coordinate a group of robots
without a central supervisor, by using only local interactions between the
robots. When this decentralized approach is used much of the communication
overhead (characteristic to centralized systems) is saved, the hardware of the
robots can be fairly simple, and better modularity is achieved. A properly
designed system should be readily scalable, achieving reliability and robustness through redundancy.

\noindent \textbf{Multi-Agent Robotics and Swarm Robotics}.
Significant research effort has been invested during the last few years in
design and simulation of multi-agent robotics and intelligent swarm systems,
e.g.
\cite{Mastellone1,DeLoach1,AMAI_Intro}.


Swarm based robotic systems can generally be defined as highly decentralized
collectives, i.e. groups of extremely simple robotic agents, with limited
communication, computation and sensing abilities, designed to be deployed together in order to accomplish various tasks.

Tasks that have been of particular interest to researchers in recent years include synergetic
mission planning \cite{RoboCup2007,survey1}, patrolling \cite{Agmon1,Agmon2}, fault-tolerant cooperation \cite{Kraus1,Work1}, network security \cite{Rehak1}, swarm control \cite{Connaughton1,survey45}, design of mission plans \cite{Manisterski1,survey42}, role assignment \cite{Boutilier1,Zheng1,survey22}, multi-robot path planning \cite{Sawhney1,Agmon2,survey69}, traffic control
\cite{Agogino1,survey55}, formation generation \cite{Bhatt1,formation_Mataric,bee_dance},
formation keeping \cite{Bendjilali1,survey13}, exploration and mapping
\cite{Sariel1,RoboCup2007_SLAM,exploration_dudek}, target tracking \cite{Harmatia1,survey53} and
distributed search, intruder detection and surveillance \cite{Hollinger1,survey39}.

Unfortunately, the mathematical / geometrical theory
of such multi agent systems is far from being satisfactory, as
pointed out in~\cite{Peleg07,Olfati-Saber06,BonabeauBook99,Intro5} and many
other papers.

\noindent \textbf{Multi Robotics in Dynamic Environments}.
The vast majority of the works mentioned above discuss challenges involving a multi agent system operating on static domains. Such models, however, are often too limited to capture ``real world'' problems which, in many cases, involve external element, which may influence their environment, activities and goals. Designing robotic agents that can operate in such environments presents a variety of mathematical challenges.

The main difference between algorithms designed for static environments and algorithms designed to work in dynamic environments is the fact that the agents' knowledge base (either central or decentralized) becomes unreliable, due to the changes that take place in the environment. Task allocation, cellular decomposition, domain learning and other approaches often used by multi agents systems --- all become impractical, at least to some extent. Hence, the agents' behavior must ensure that the agents generate a desired effect, regardless  the changing environment.

One example is the use of multi agents for distributed search. While many works discuss search after ``idle targets'',
recent works considered dynamic targets, meaning targets which while being searched for by the searching robots, respond by performing various evasive
maneuvers intended to prevent their interception.
This problem, dating back to World War II operations research (see e.g. \cite{WWII1,WWII2}), requires the robotic agents to cope with a search area that expands while scanned. The first documented example for search in dynamic domains discussed a planar search problem,
considering the scanning of a corridor between parallel borders. This problem was solved in~\cite{Koopman} in order to determine
optimal strategies for aircraft searching for moving ships in a channel.

A similar problem was presented in \cite{UAV_UCLA}, where a system consisting of
a swarm of UAVs (Unmanned Air Vehicles) was designed to search for one or more
``smart targets'' (representing for example enemy units, or alternatively
a lost friendly unit which should be found and rescued). In this problem
the objective of the UAVs is to find the targets in the shortest time possible. While the swarm comprises
relatively simple UAVs, lacking prior knowledge of the initial positions of the
targets, the targets are assumed to be adversarial and equipped with strong sensors, capable of telling the
locations of the UAVs from very long distances.
The search strategy suggested in \cite{UAV_UCLA} defines
\emph{flying patterns} for the UAVs to follow, designed for scanning
the (rectangular) area in such a way that the targets cannot
re-enter areas which were already scanned by the swarm without
being detected. This problem was further discussed in \cite{UAV-ROBOTICA}, where an improved decentralized search strategy was discussed,
demonstrating nearly optimal completion time, compared to the theoretical optimum achievable by any search algorithm.

\noindent \textbf{Collaborative Coverage of Expanding Domains}.
In this paper we shall examine a problem in which a group of ant-like robotic agents must cover an unknown region in the grid, that possibly expands over time. This problem is also strongly related to the problem of distributed search after mobile and evading target(s) \cite{Koenig-movingtargets,Koenig-pursuits-complexity,UAV-ROBOTICA} or the problems discussed under the names of ``Cops and Robbers'' or ``Lions and Men'' pursuits \cite{Lions-Isaacs,Lions-Isler,Lions-Goldstein,Lions-Flynn}.

We analyze such issues using the results presented in~\cite{CC08,dCC,ICARA-upper}, concerning the \emph{Cooperative Cleaners} problem, a problem that assumes a regular grid of connected `pixels' / `tiles' / `squares' / `rooms', part of which are `dirty', the `dirty' pixels forming a
connected region of the grid. On this dirty grid region several agents move,
each having the ability to `clean' the place (the `room', `tile', `pixel' or `square') it
is located in. In the dynamic variant of this problem a deterministic evolution of the
environment in assumed, simulating a spreading \emph{contamination} (or spreading
\emph{fire}).
In the spirit of~\cite{Intro6} we consider simple robots with only a bounded
amount of memory (i.e.~\emph{finite-state-machines}).

First, we discuss the collaborative coverage of static grids. We demonstrate that the best completion time known to date ($O(n^{2})$, achievable for example using the \emph{LRTA*} search algorithm \cite{LRTASTAR}) can be improved to guarantee grid coverage in $O(\frac{1}{k} n^{1.5} + n)$.

Later, we discuss the problem of covering an expanding domain, namely~---~a region in which ``covered'' tiles that are adjacent to ``uncovered'' tiles become ``uncovered'' every once in a while. Note that the grid is infinite, namely although initial size of the region is $n$, it can become much greater over time. We show that using any conceivable algorithm, and using as sophisticated and potent robotic agents as possible, the minimal number of robots below which covering such a region is impossible equals $\Omega({\sqrt{n}})$.
We then show that when the region expands sufficiently slow, specifically --- every $O(\frac{c_{0}}{\gamma_{1}})$ time steps (where $c_{0}$ is the circumference of the region and where $\gamma_{1}$ is a geometric property of the region, which ranges between $O(1)$ and $O(\ln n)$), a group of $\Theta(\sqrt{n})$ robots can successfully cover the region. Furthermore, we demonstrate that in this case a cover time of $O(n^{2} \ln n)$ can be guaranteed.
%

These results are the first analytic results ever concerning the complexity of the number of robots required to cover an expanding grid, as well as for the time such a coverage requires.

\section{Related Work}
\label{Section-Related}

In general, most of the techniques used for the task of a distributed coverage
use some sort of cellular decomposition. For example,
in~\cite{primitive_static_cleaning} the area to be covered is divided between
the agents based on their relative locations. In~\cite{butler1} a different
decomposition method is being used, which is analytically shown to guarantee a
complete coverage of the area. Another interesting work is presented
in~\cite{choset2}, discussing two methods for cooperative coverage (one probabilistic and the other based on an exact cellular
decomposition). All of the works mentioned above, however, rely on the
assumption that the cellular decomposition of the area is possible. This in
turn, requires the use of memory resources, used for storing the dynamic map
generated, the boundaries of the cells, etc'. As the initial size and geometric
features of the area are generally not assumed to be known in advance, agents
equipped with merely a constant amount of memory will not be able to use such
algorithms.

Surprisingly, while some existing works concerning distributed (and
decentralized) coverage present analytic proofs for the ability of the system to guarantee the completion of the task
(for example, in~\cite{choset2,butler1,batalin1}), most of them
lack analytic bounds for the coverage time (although in many cases an
extensive amount of empirical results of this nature are made available by extensive simulations).

An interesting work discussing a decentralized coverage of terrains is presented in \cite{Zheng2}. This work examines domains with non-uniform traversability. Completion times are given for the proposed algorithm, which is a generalization of the forest search algorithm \cite{Zheng05multi-robotforest}. In this work, though, the region to be searched is assumed to be known in advance --- a crucial assumption for the search algorithm, which relies on a cell-decomposition procedure.

A search for analytic results concerning the completion time of ant-robots covering an
area in the grid revealed only a handful of works. The main result in this regard is
that of~\cite{Svennebring1,similar_terrain_coverage}, where a swarm of ant-like robots is
used for repeatedly covering an unknown area, using a real time search method
called \emph{node counting}. By using this method, the robots are shown to be able
to efficiently perform such a coverage mission (using integer markers that are placed on the graph's nodes), and analytic bounds for the coverage
time are discussed. Based on a more general result for strongly connected undirected
graphs shown in~\cite{Koenig2,Szymanski1}, the cover time of teams of ant robots (of a given size) that use node counting is shown to be $t_{k}(n) = O(n^{\sqrt{n}})$, when $t_{k}(n)$ the cover time of a region of size $n$ using $k$ robots.
It should be mentioned though, that in~\cite{Svennebring1} the authors clearly
state that it is their belief that the coverage time for robots using node
counting in grids is much smaller. This evaluation is also demonstrated
experimentally. However, no analytic evidence for this was available thus far.

Another algorithm to be mentioned in this scope is the \emph{LRTA*} search algorithm. This
algorithm was first introduced in \cite{LRTASTAR} and its multi-robotics variant is shown in \cite{Koenig2} to guarantee cover time of undirected connected graphs in polynomial time. Specifically, on grids this algorithm is shown to guarantee coverage in $O(n^2)$ time (again, using integer markers).

\emph{Vertex-Ant-Walk}, a variant of the node counting algorithm is presented in \cite{GraphSearch1}, is shown to achieve a coverage time of $O(n \delta_{G})$, where $\delta_{G}$ is the graph's diameter. Specifically, the cover time of regions in the grid is expected to be $O(n^{2})$ (however for various ``round'' regions, a cover time of approximately $O(n^{1.5})$ can be achieved). This work is based on a previous work in which a cover time of $O(n^{2} \delta_{G})$ was demonstrated \cite{Thrun-coverage}.

Another work called \emph{Exploration as Graph Construction}, provides a coverage of degree bounded graphs in $O(n^{2})$ time, can be found in \cite{Exploration-Dudek}. Here a group of limited ant robots explore an unknown graph using special ``markers''.

Interestingly, a similar performance can be obtained by using the simplest algorithm for multi robots navigation, namely~---~random walk. Although in general undirected graphs a group of $k$ random walking robots may require up to $O(n^{3})$ time, in degree bounded undirected graphs such robots would achieve a much faster covering, and more precisely, $O \left( \frac{|E|^{2} log^{3}n}{k^{2}} \right)$ \cite{exploration_broder}. For regular graphs or degree bounded planar graphs a coverage time of $O(n^{2})$ can be achieved \cite{random-walkers-Lovasz}, although in such case there is also a lower bound for the coverage time, which equals $\Omega(n (\log n)^{2})$ \cite{schramm-covertime-planar}.

We next show that the problem of collaborative coverage in static grid domains can be completed in $O(\frac{1}{k} n^{1.5} + n)$ time and that collaborative coverage of dynamic grid domains can be achieved in $O(n^{2} \ln n)$.

\section{The Dynamic Cooperative Cleaners Problem}
\label{Section-CleaningProblem}

Following is a short summary of the \emph{Cooperative Cleaners} problem, as appears in \cite{CC08} (static variant) and \cite{dCC,ICARA-upper} (dynamic variant).

We shall assume that the time is discrete. Let the undirected graph $G(V,E)$ denote a two
dimensional integer grid $\mbox{\bf Z}^{2}$, whose vertices (or
``\emph{tiles}'') have a binary property called `\emph{contamination}'. Let
$cont_{t}(v)$ state the contamination state of the tile $v$ at time $t$, taking
either the value ``\emph{on}'' or ``\emph{off}''.
Let $F_{t}$ be the contaminated sub-graph of $G$ at time $t$,
i.e.~: $F_{t} = \left\{v \in G \ | \ cont_{t}(v) = on \right\}$.
We assume that $F_{0}$ is a single connected component. Our algorithm will
preserve this property along its evolution.

Let a group of $k$ robots that can move on the grid $G$ (moving from a tile
to its neighbor in one time step) be placed at time $t_{0}$ on $F_{0}$, at point $p_{0} \in F_{t}$.
Each robot is equipped with a sensor capable of telling the
contamination status of all tiles in the digital sphere of diameter 7,
surrounding the robot (namely, in all the tiles that their Manhattan distance from the robot is equal or smaller than 3. See an illustration in Figure \ref{fig.digitalsphere}). A robot is also aware of other robots which are
located in these tiles, and all the robots agree on a common direction. Each
tile may contain any number of robots simultaneously.
Each robot is equipped with a memory of size $O(1)$ bits\footnote{For counting purposes the agents must be equipped with counters that can store the number of agents in their immediate vicinity. This can of course be implemented using $O(\log k)$ memory. However, throughout the proof of Lemma 5 in \cite{CC08} it is shown that the maximal number of agents that may simultaneously reside in the same tile at any given moment is upper bounded by $O(1)$. Therefore, counting the agents in the immediate vicinity can be done using counters of $O(1)$ bits.}.
When a robot moves to a tile $v$, it has the possibility of cleaning this tile
(i.e.~causing $cont(v)$ to become \emph{off}.
The robots do not have any prior knowledge of the shape or size of the sub-graph $F_{0}$ except that it is a single and simply connected component.

\setlength{\unitlength}{0.25cm}
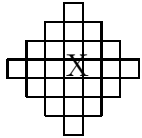
\begin{figure}[htbp]
\begin{center}
\begin{picture}(8,7)
\multiput(0,3)(1,0){8}{\line(0,1){1}} 
\multiput(1,2)(1,0){6}{\line(0,1){1}} 
\multiput(1,4)(1,0){6}{\line(0,1){1}} 
\multiput(2,1)(1,0){4}{\line(0,1){1}} 
\multiput(2,5)(1,0){4}{\line(0,1){1}} 
\multiput(3,0)(1,0){2}{\line(0,1){1}} 
\multiput(3,6)(1,0){2}{\line(0,1){1}} 
\multiput(3,0)(0,1){8}{\line(1,0){1}} 
\multiput(2,1)(0,1){6}{\line(1,0){1}} 
\multiput(4,1)(0,1){6}{\line(1,0){1}} 
\multiput(1,2)(0,1){4}{\line(1,0){1}} 
\multiput(5,2)(0,1){4}{\line(1,0){1}} 
\multiput(0,3)(0,1){2}{\line(1,0){1}} 
\multiput(6,3)(0,1){2}{\line(1,0){1}} 
\put(3.1,3.1){X}
\end{picture}
\end{center}
\caption[Digital sphere of diameter 7] %
{An illustration of a digital sphere of diameter 7, placed around a robot.} %
\label{fig.digitalsphere}
\end{figure}

\begin{figure}[htbp]
\begin{center}
\epsfig{figure=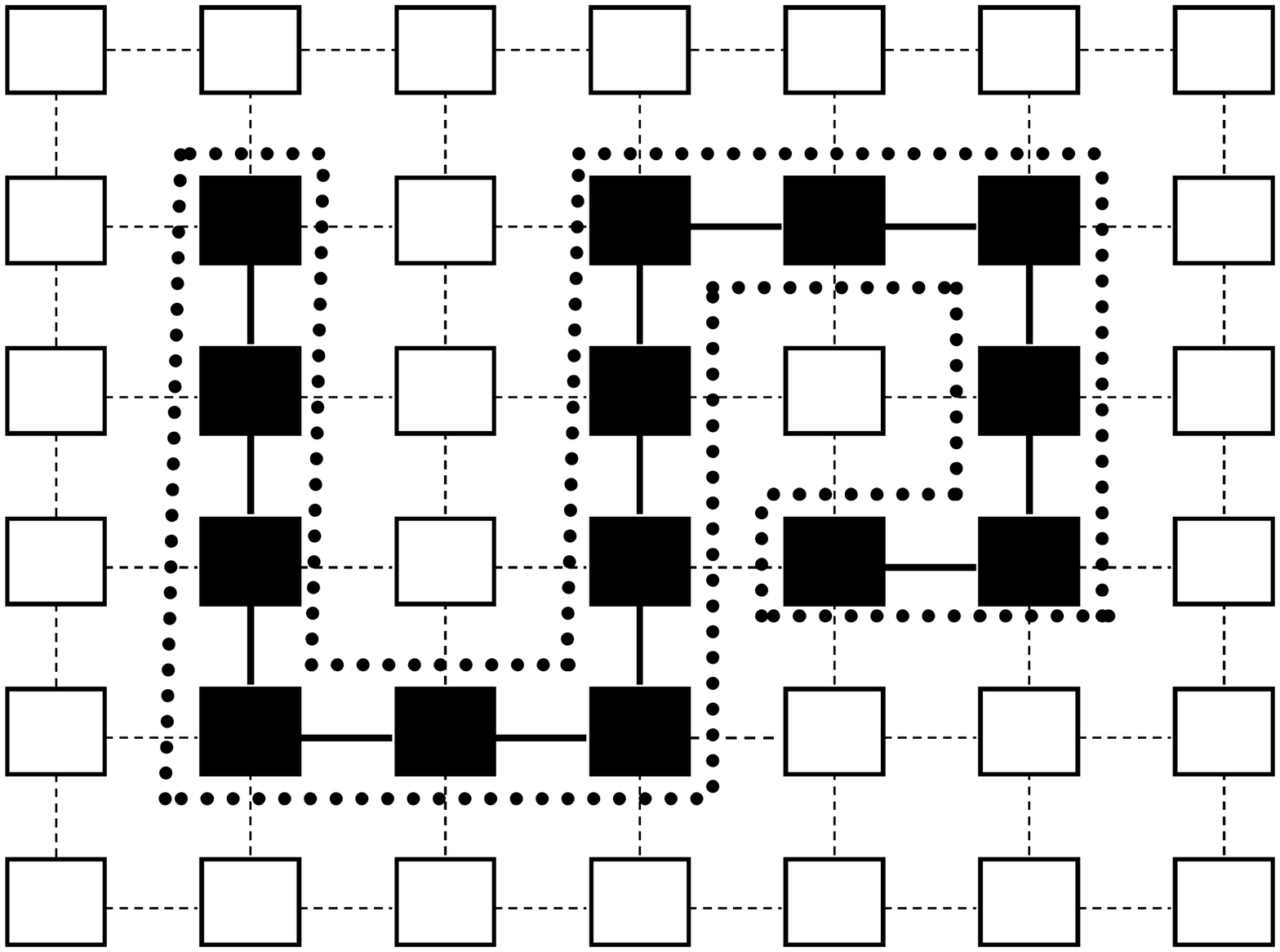,width=0.8in,angle=-90} %
\epsfig{figure=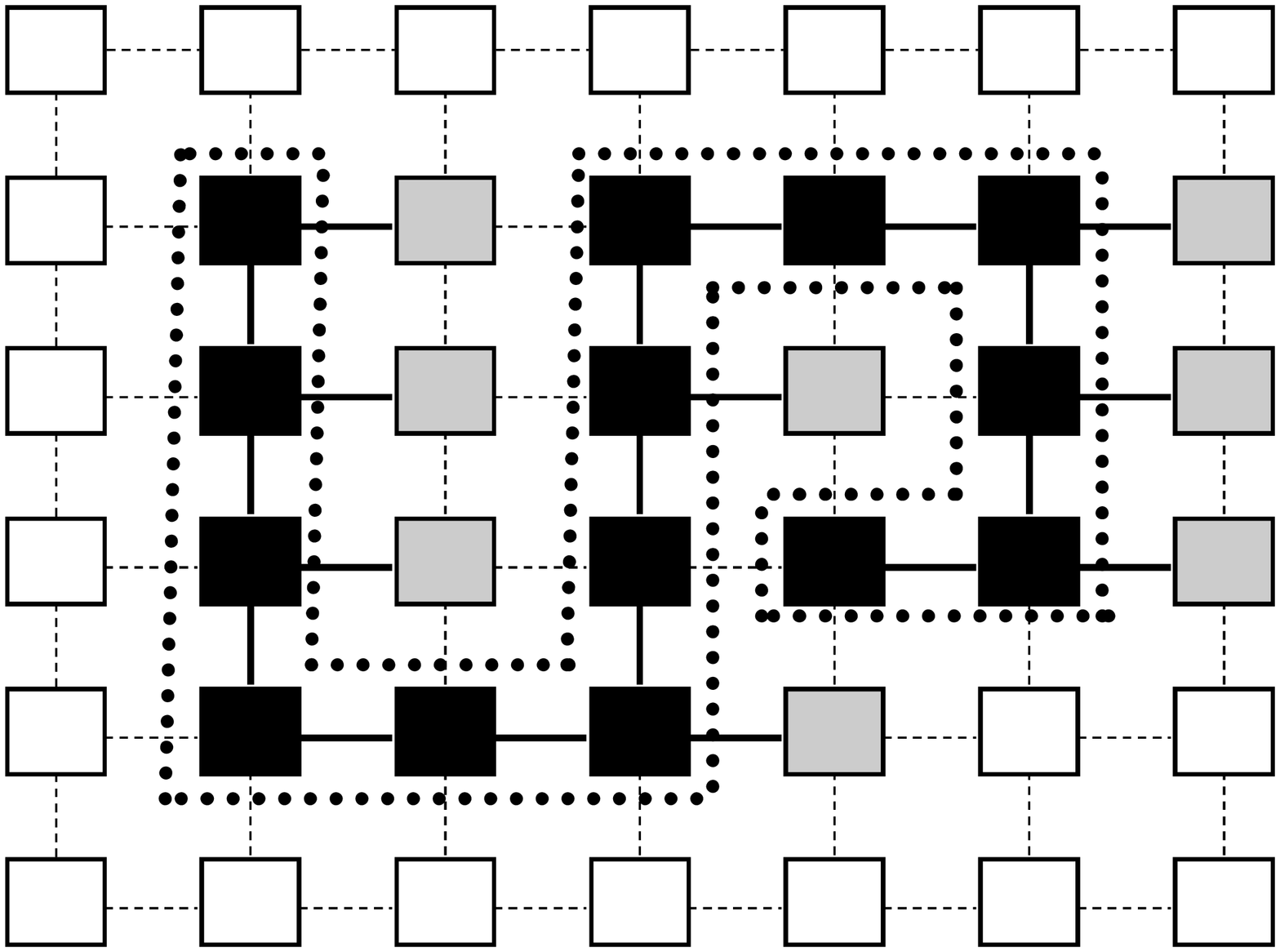,width=0.8in,angle=-90} %
\epsfig{figure=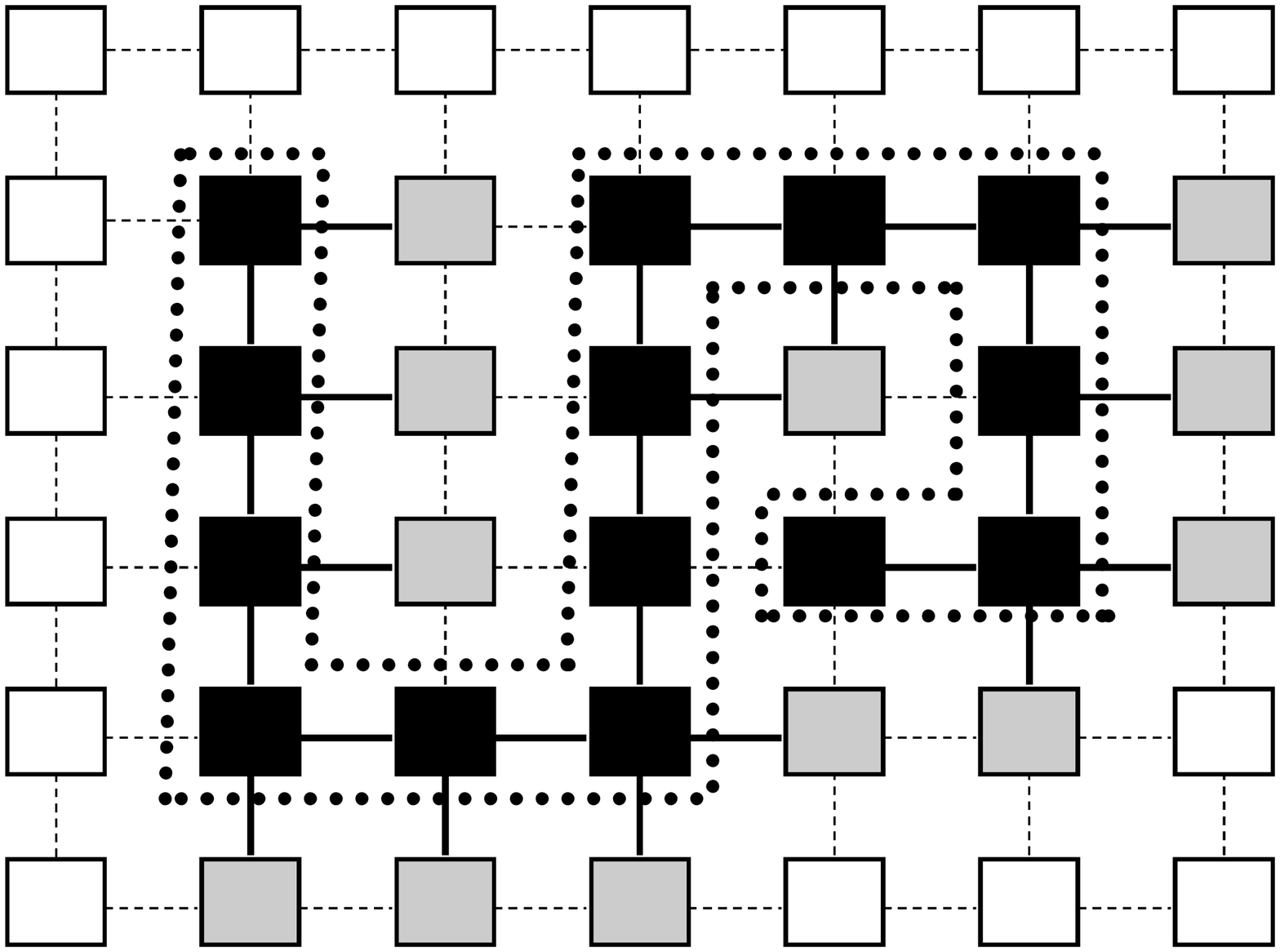,width=0.8in,angle=-90} %
\epsfig{figure=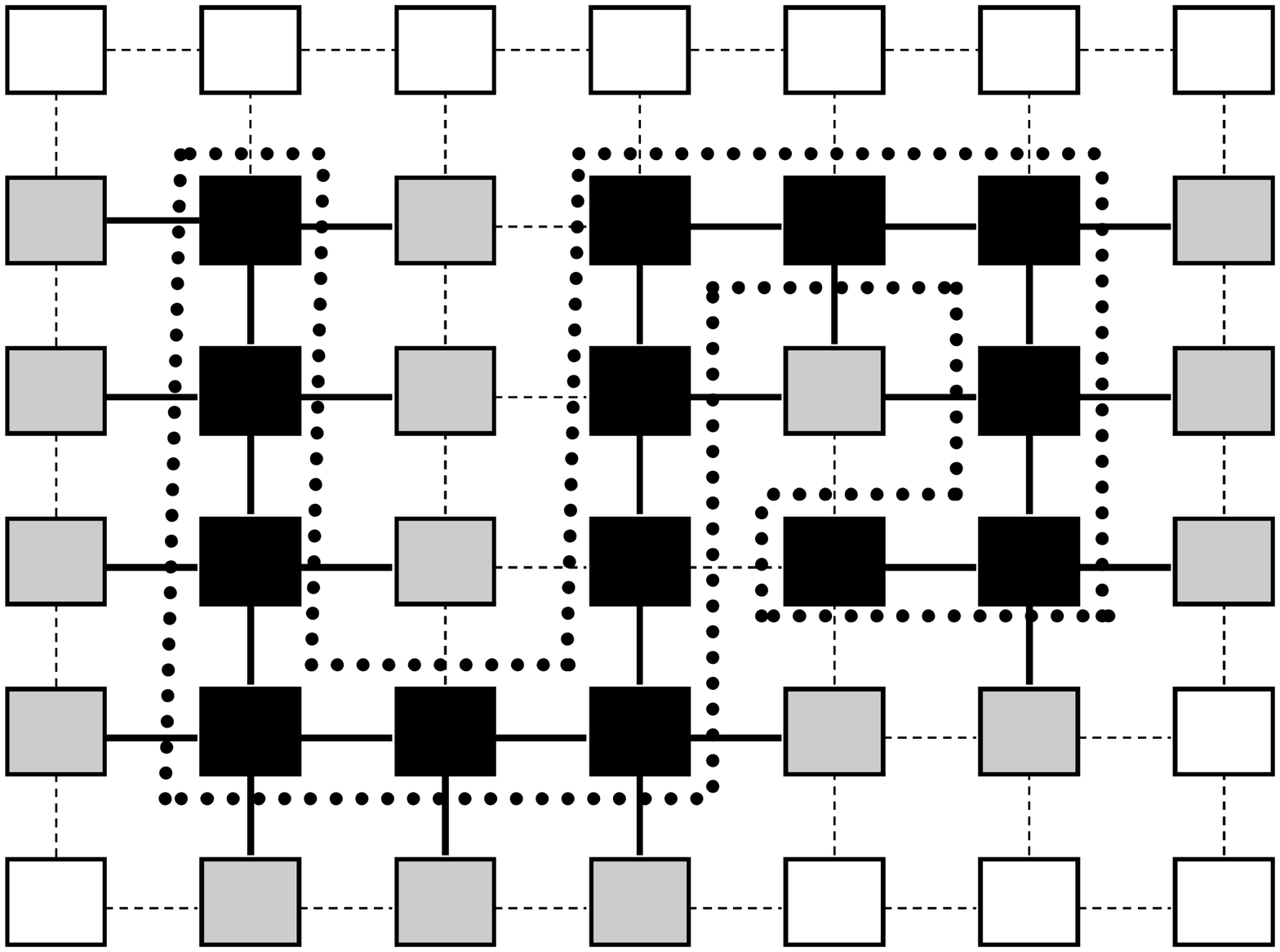,width=0.8in,angle=-90} %
\epsfig{figure=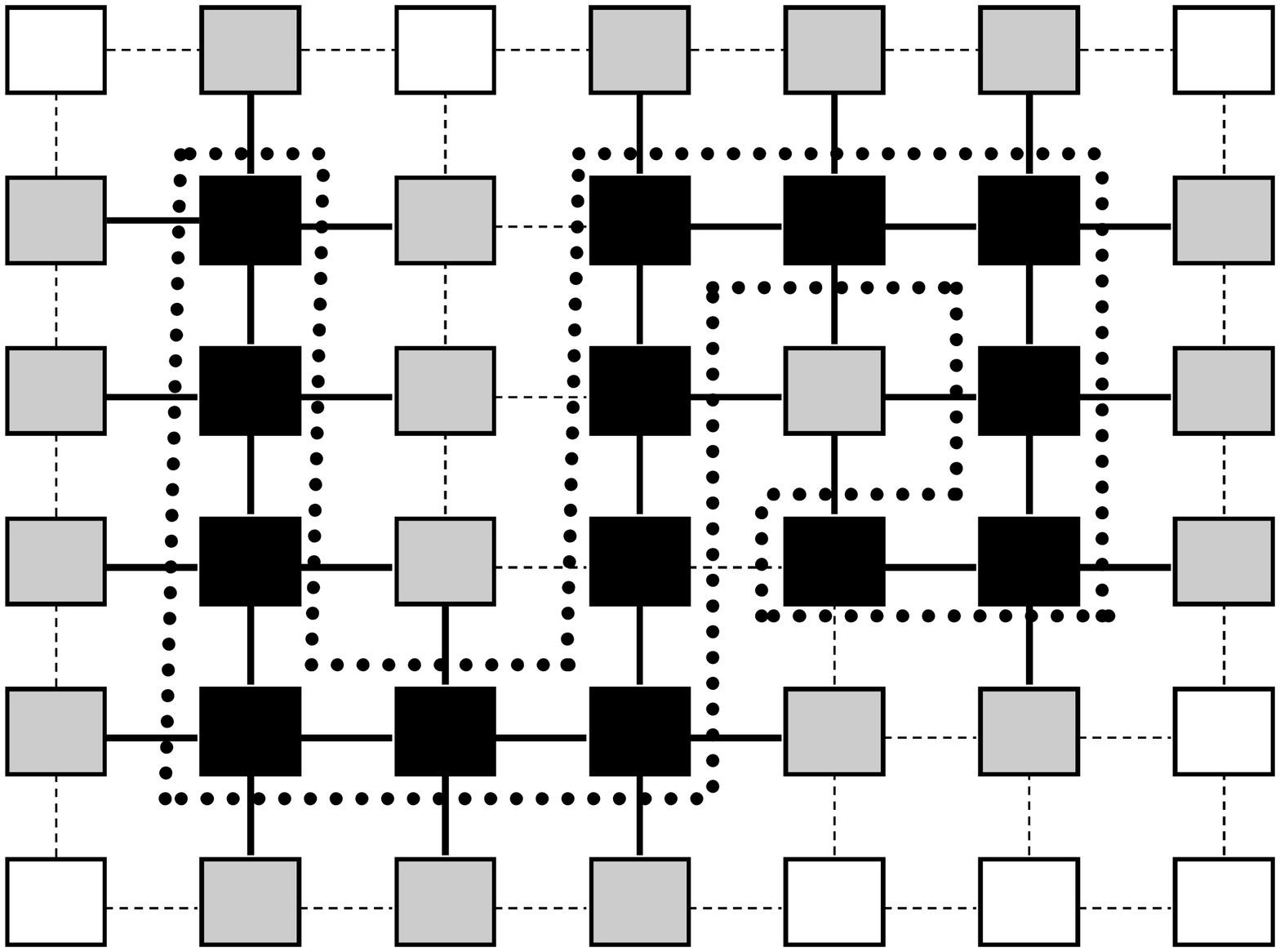,width=0.8in,angle=-90} %
\epsfig{figure=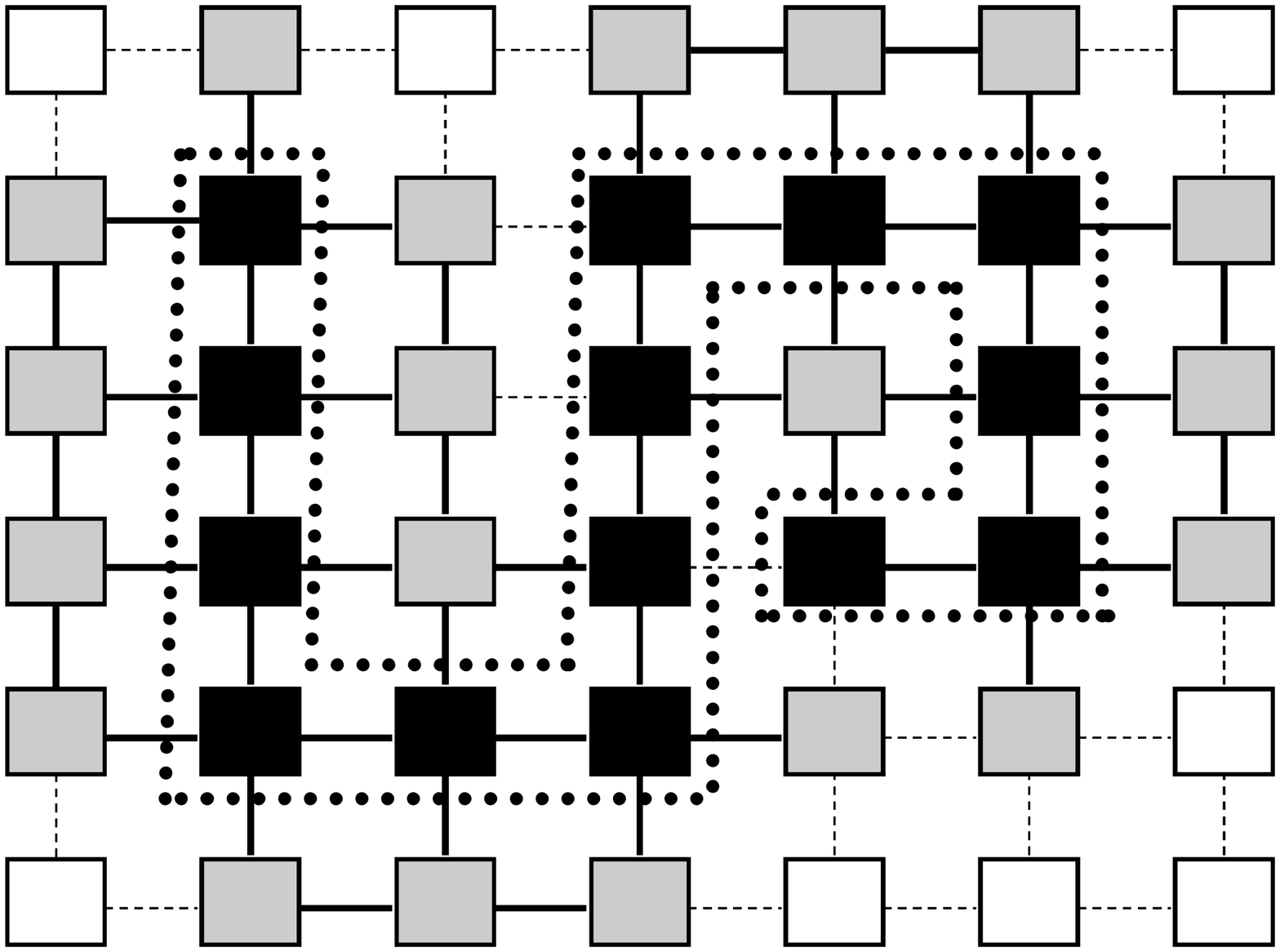,width=0.8in,angle=-90} %
\epsfig{figure=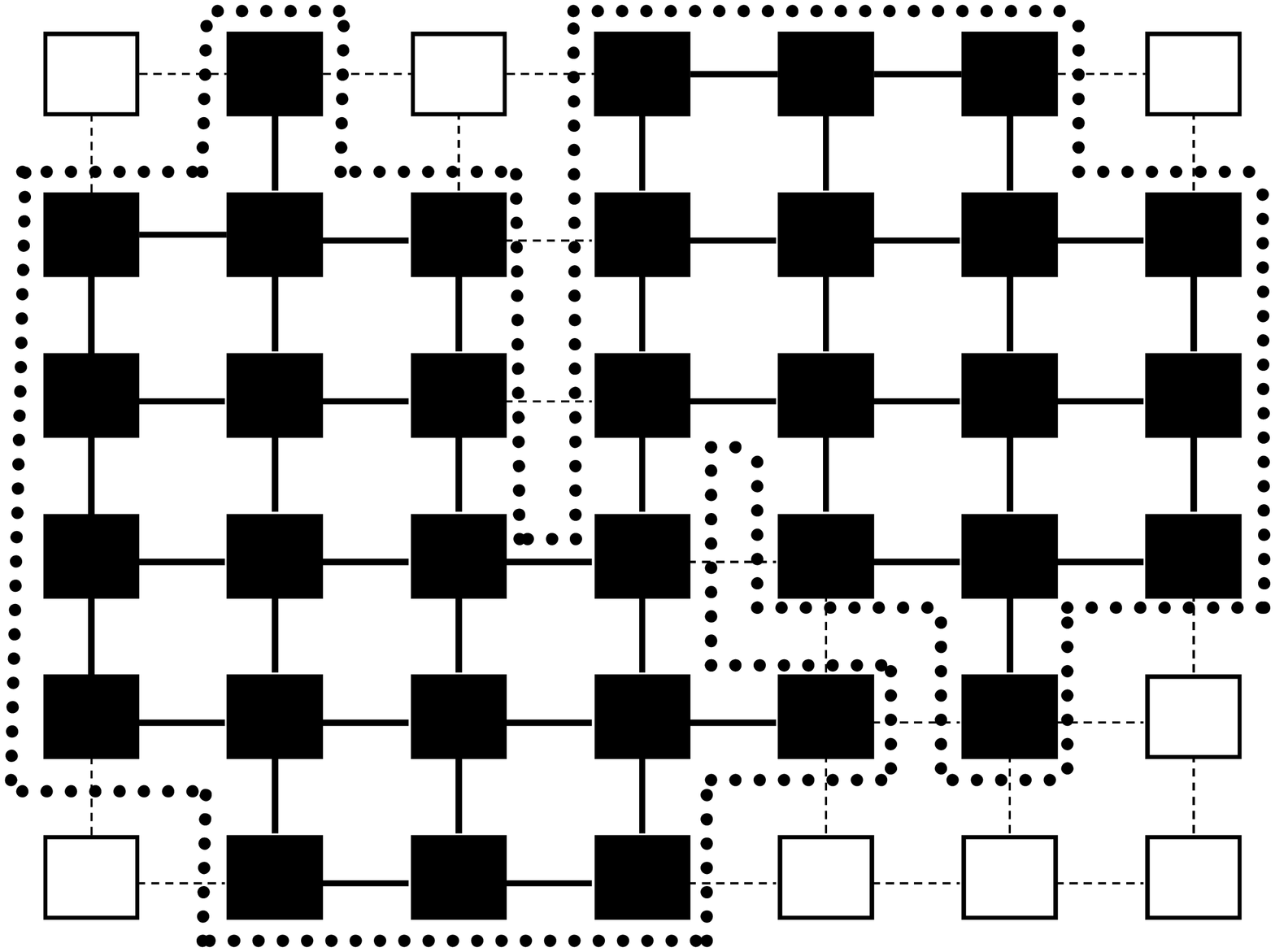,width=0.8in,angle=-90} %
\caption{A demonstration of the barrier expansion process as a result of a contamination spread.}%
\label{Figure-spreading2}
\end{center}
\end{figure}

The contaminated region $F_{t}$ is assumed to be surrounded at its boundary by a
rubber-like elastic barrier, dynamically reshaping itself to fit the evolution
of the contaminated region over time.
This barrier is intended to guarantee the preservation of the simple
connectivity of $F_{t}$, crucial for the operation of the robots, due to their limited memory.
When a robot cleans a contaminated tile, the barrier retreats, in
order to fit the void previously occupied by the cleaned tile.
Every $d$ time steps, the contamination spreads. That is, if $t = nd$ for some
positive integer $n$, then~:
\[\forall v \in F_{t} \ \forall u \in 4-Neighbors(v) \ , \ cont_{t+1}(u) = on\]
Here, the term $4-Neighbors(v)$ simply means the four tiles adjacent to tile $v$ (namely, the tiles whose Manhattan distance from $v$ equals 1).
While the contamination spreads, the elastic barrier stretches while preserving the simple-connectivity of the region, as demonstrated in Figure~\ref{Figure-spreading2}.
For the robots who travel along the tiles of $F$, the barrier signals the
boundary of the contaminated region.

%
%
%

The robots' goal is to clean $G$ by eliminating the contamination entirely.

It is important to note that no central control is allowed, and that the system is fully decentralized (i.e.~all robots are identical and no explicit
communication between the robots is allowed).
An important advantage of this approach, in addition to the simplicity of the
robots, is fault-tolerance --- even if almost all the robots die and evaporate
before completion, the remaining ones will eventually complete the mission, if
possible.

\noindent \textbf{A Survey of Previous Results}.
The cooperative cleaners problem was previously studied in \cite{CC08} (static version)
and \cite{dCC,ICARA-upper} (dynamic version). A cleaning algorithm called
\textbf{SWEEP} was proposed (used by a decentralized group of simple mobile robots,
for exploring and cleaning an unknown ``contaminated'' sub-grid $F$, expanding
every $d$ time steps) and its performance analyzed.

The \textbf{SWEEP} algorithm is based in a constant traversal of the contaminated region, preserving the connectivity of the region while cleaning all \emph{non critical points} --- points which when cleaned disconnect the contaminated region. Using this algorithm the agents are guaranteed to stop only upon completing their mission. The algorithm can be implemented using only local knowledge, and local interactions by immediately adjacent agents.
At each time step, each agent cleans its current location (assuming it is not
a critical point), and moves according to a local movement rule, creating the
effect of a clockwise ``sweeping'' traversal along the boundary of the contaminated region. As a result, the agents ``peel'' layers from the region, while preserving its connectivity, until
the region is cleaned entirely. An illustration of two agents working according to the protocol can be seen in
Figure~\ref{figure_example_protocol}.

\setlength{\unitlength}{0.3cm}
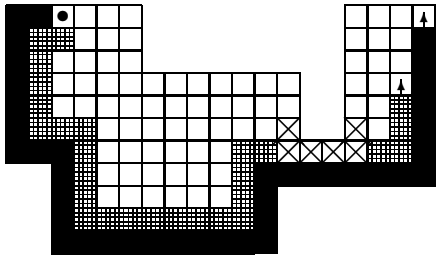
\begin{figure}[htbp]
\begin{center}
\begin{picture}(19,12)(0,-1)
\DefineSquareEmpty %
\DefineSquareBarsA %
\DefineSquareFull %
\DefineSquareWithX
\multiput(0,4)(1,0){6}{\multiput(0,0)(0,1){7}{\usebox{\SquareEmpty}}}
\multiput(2,0)(1,0){9}{\multiput(0,0)(0,1){8}{\usebox{\SquareEmpty}}}
\multiput(11,3)(1,0){2}{\multiput(0,0)(0,1){5}{\usebox{\SquareEmpty}}}
\multiput(13,3)(1,0){2}{\multiput(0,0)(0,1){2}{\usebox{\SquareEmpty}}}
\multiput(15,3)(1,0){4}{\multiput(0,0)(0,1){8}{\usebox{\SquareEmpty}}}
\multiput(11.8,3.8)(1,0){4}{\line(1,1){1.2}}
\multiput(12,5)(1,0){4}{\line(1,-1){1.2}}
\multiput(11.8,4.8)(3,0){2}{\line(1,1){1.2}}
\multiput(12,6)(3,0){2}{\line(1,-1){1.2}}
\multiput(0,4)(18,-1){2}{\multiput(0,0)(0,1){7}{\usebox{\SquareFull}}}
\multiput(1,4)(0,6){2}{\usebox{\SquareFull}}
\multiput(2,4)(0,-1){5}{\usebox{\SquareFull}}
\multiput(2,0)(1,0){9}{\usebox{\SquareFull}}
\multiput(11,0)(0,1){4}{\usebox{\SquareFull}}
\multiput(11,3)(1,0){8}{\usebox{\SquareFull}}
\multiput(1,5)(2,-4){2}{\multiput(0,0)(0,1){5}{\usebox{\SquareBarsA}}}
\multiput(2,5)(0,4){2}{\usebox{\SquareBarsA}}
\multiput(3,1)(1,0){7}{\usebox{\SquareBarsA}}
\multiput(10,1)(0,1){4}{\usebox{\SquareBarsA}}
\multiput(10,4)(6,0){2}{\multiput(0,0)(1,0){2}{\usebox{\SquareBarsA}}}
\multiput(17,5)(0,1){2}{\usebox{\SquareBarsA}}
\put(2.5,10.5){\circle*{0.5}} %
\multiput(17.5,7)(1,3){2}{\vector(0,1){0.7}}
\end{picture}
\end{center}
\caption[An example of the \textbf{SWEEP} protocol] %
{An example of two agents using the \textbf{SWEEP} protocol, at time step 40
(with contamination spreading speed $d > 40$). All the tiles presented were
contaminated at time 0. The black dot denotes the starting point of the agents.
The X's mark the \CritPoints{} which are not cleaned. The black tiles are the
tiles cleaned by the first agent. The second layer of marked tiles represent
the tiles cleaned by the second agent.} %
\label{figure_example_protocol}
\end{figure}

In order to formally describe the \textbf{SWEEP} algorithm, we should first define several additional terms.
Let $\tau(t) = \big(\tau_{1}(t), \tau_{2}(t), \ldots, \tau_{k}(t)\big)$ denote the locations of the $k$ agents at
time $t$.
In addition, let $\tilde{\tau_{i}}(t)$ denote the ``previous location'' of agent $i$. Namely, the last tile that agent $i$ had been at, which is different than
$\tau_{i}(t)$. This is formally defined as~:
\[
\tilde{\tau_{i}}(t) \triangleq \tau_{i}(x) \emph{ s.t. } x = \max\{j \in \mathbb{N} \ | \ j < t \emph{ and } \tau_{i}(j) \neq \tau_{i}(t)\}\]

The term $\partial F$ denotes the boundary of $F$, defined via~:
\begin{displaymath}
\partial F = \{v \ | \ v \in F \ \wedge \ 8-Neighbors(v) \ \cap \
(G \ \setminus \ F) \ \neq \ \emptyset\}
\end{displaymath}

The term `\emph{rightmost}' can now be defined as follows~:
\begin{itemize}
  \item If $t = 0$ then select the tile as instructed in Figure~\ref{figure_rightmost}.
  \item If $\tilde{\tau_{i}}(t) \in \partial F_{t}$ then starting from $\tilde{\tau_{i}}(t)$ (namely, the previous boundary tile that the agent
had been in) scan the \emph{four neighbors} of $\tau_{i}(t)$ in a clockwise order until a boundary
tile (excluding $\tilde{\tau_{i}}(t)$) is found.
  \item If not $\tilde{\tau_{i}}(t) \in \partial F_{t}$ then starting from $\tilde{\tau_{i}}(t)$ scan the \emph{four neighbors} of $\tau_{i}(t)$ in a clockwise order until the second boundary tile is found.
\end{itemize}

\setlength{\unitlength}{0.25cm}
\begin{figure}[htbp]
\begin{center}
\begin{tabular}{cccccccc}
\begin{picture}(4,4)(-0.5,-0.5) \DefineSquareEmpty
\multiput(0,0)(1,0){2}{\usebox{\SquareEmpty}} \put(0.5,0.5){\circle*{0.5}}
\put(1.5,0.5){\circle{0.5}}
\end{picture}
&
\begin{picture}(4,4)(-0.5,-0.5) \DefineSquareEmpty
\multiput(0,0)(1,0){3}{\usebox{\SquareEmpty}} \put(1.5,0.5){\circle*{0.5}}
\put(2.5,0.5){\circle{0.5}}
\end{picture}
&
\begin{picture}(4,4)(-0.5,-0.5) \DefineSquareEmpty
\multiput(0,0)(1,0){2}{\usebox{\SquareEmpty}}
\put(1,1){\usebox{\SquareEmpty}} %
\put(1.5,0.5){\circle*{0.5}} \put(1.5,1.5){\circle{0.5}}
\end{picture}
&
\begin{picture}(4,4)(-0.5,-0.5) \DefineSquareEmpty
\multiput(0,0)(1,0){2}{\usebox{\SquareEmpty}}
\put(1,1){\usebox{\SquareEmpty}} %
\put(0.5,0.5){\circle*{0.5}} \put(1.5,0.5){\circle{0.5}}
\end{picture}
&
\begin{picture}(4,4)(-0.5,-0.5) \DefineSquareEmpty
\multiput(0,0)(1,0){2}{\usebox{\SquareEmpty}}
\multiput(0,1)(0,1){2}{\usebox{\SquareEmpty}} %
\put(0.5,1.5){\circle*{0.5}} \put(0.5,2.5){\circle{0.5}}
\end{picture}
&
\begin{picture}(4,4)(-0.5,-0.5) \DefineSquareEmpty
\multiput(0,0)(0,1){3}{\usebox{\SquareEmpty}}
\put(1,1){\usebox{\SquareEmpty}} %
\put(0.5,1.5){\circle*{0.5}} \put(0.5,2.5){\circle{0.5}}
\end{picture}
&
\begin{picture}(4,4)(-0.5,-0.5) \DefineSquareEmpty
\multiput(0,0)(0,1){3}{\usebox{\SquareEmpty}}
\put(1,1){\usebox{\SquareEmpty}} %
\put(1.5,1.5){\circle*{0.5}} \put(0.5,1.5){\circle{0.5}}
\end{picture}
&
\begin{picture}(4,4)(-0.5,-0.5) \DefineSquareEmpty
\multiput(0,0)(1,0){2}{\multiput(0,0)(0,1){2}{\usebox{\SquareEmpty}}} %
\put(0.5,0.5){\circle*{0.5}} \put(0.5,1.5){\circle{0.5}}
\end{picture}
\\   
\begin{picture}(4,4)(-0.5,-0.5) \DefineSquareEmpty
\multiput(0,0)(1,1){2}{\multiput(0,0)(0,1){2}{\usebox{\SquareEmpty}}} %
\put(0.5,1.5){\circle*{0.5}} \put(1.5,1.5){\circle{0.5}}
\end{picture}
&
\begin{picture}(4,4)(-0.5,-0.5) \DefineSquareEmpty
\multiput(0,0)(1,0){2}{\multiput(0,0)(0,1){2}{\usebox{\SquareEmpty}}}
\put(0,2){\usebox{\SquareEmpty}} %
\put(0.5,1.5){\circle*{0.5}} \put(0.5,2.5){\circle{0.5}}
\end{picture}
&
\begin{picture}(4,4)(-0.5,-0.5) \DefineSquareEmpty
\multiput(0,0)(1,0){2}{\multiput(0,0)(0,1){2}{\usebox{\SquareEmpty}}}
\put(0,2){\usebox{\SquareEmpty}} %
\put(1.5,1.5){\circle*{0.5}} \put(1.5,0.5){\circle{0.5}}
\end{picture}
&
\begin{picture}(4,4)(-0.5,-0.5) \DefineSquareEmpty
\multiput(0,0)(0,2){2}{\multiput(0,0)(1,0){2}{\usebox{\SquareEmpty}}}
\put(0,1){\usebox{\SquareEmpty}} %
\put(0.5,1.5){\circle*{0.5}} \put(0.5,2.5){\circle{0.5}}
\end{picture}
&
\begin{picture}(4,4)(-0.5,-0.5) \DefineSquareEmpty
\multiput(0,0)(1,0){3}{\usebox{\SquareEmpty}}
\multiput(1,1)(0,1){2}{\usebox{\SquareEmpty}} %
\put(1.5,1.5){\circle*{0.5}} \put(1.5,2.5){\circle{0.5}}
\end{picture}
&
\begin{picture}(4,4)(-0.5,-0.5) \DefineSquareEmpty
\multiput(1,0)(1,0){2}{\multiput(0,0)(-1,1){2}{\usebox{\SquareEmpty}}}
\put(1,2){\usebox{\SquareEmpty}} %
\put(1.5,1.5){\circle*{0.5}} \put(1.5,2.5){\circle{0.5}}
\end{picture}
&
\begin{picture}(4,4)(-0.5,-0.5) \DefineSquareEmpty
\multiput(1,0)(1,0){2}{\multiput(0,0)(-1,2){2}{\usebox{\SquareEmpty}}}
\put(1,1){\usebox{\SquareEmpty}} %
\put(1.5,1.5){\circle*{0.5}} \put(1.5,2.5){\circle{0.5}}
\end{picture}
&
\begin{picture}(4,4)(-0.5,-0.5) \DefineSquareEmpty
\multiput(1,0)(0,1){3}{\usebox{\SquareEmpty}}
\multiput(0,1)(1,0){3}{\usebox{\SquareEmpty}} %
\put(1.5,1.5){\circle*{0.5}} \put(1.5,2.5){\circle{0.5}}
\end{picture}
\\   
\begin{picture}(4,4)(-0.5,-0.5) \DefineSquareEmpty
\multiput(1,1)(1,0){2}{\multiput(0,0)(-1,1){2}{\usebox{\SquareEmpty}}}
\put(1,0){\usebox{\SquareEmpty}} %
\put(1.5,1.5){\circle*{0.5}} \put(1.5,2.5){\circle{0.5}}
\end{picture}
&
\begin{picture}(4,4)(-0.5,-0.5) \DefineSquareEmpty
\multiput(0,1)(1,0){2}{\multiput(0,0)(1,1){2}{\usebox{\SquareEmpty}}}
\put(0,0){\usebox{\SquareEmpty}} %
\put(1.5,1.5){\circle*{0.5}} \put(1.5,2.5){\circle{0.5}}
\end{picture}
&
\begin{picture}(4,4)(-0.5,-0.5) \DefineSquareEmpty
\multiput(0,0)(1,0){2}{\multiput(0,0)(0,1){3}{\usebox{\SquareEmpty}}} %
\put(0.5,1.5){\circle*{0.5}} \put(0.5,2.5){\circle{0.5}}
\end{picture}
&
\begin{picture}(4,4)(-0.5,-0.5) \DefineSquareEmpty
\multiput(1,0)(1,0){2}{\multiput(0,0)(0,1){2}{\usebox{\SquareEmpty}}}
\multiput(0,2)(1,0){2}{\usebox{\SquareEmpty}} %
\put(1.5,1.5){\circle*{0.5}} \put(1.5,2.5){\circle{0.5}}
\end{picture}
&
\begin{picture}(4,4)(-0.5,-0.5) \DefineSquareEmpty
\multiput(1,0)(1,0){2}{\multiput(0,0)(0,1){2}{\usebox{\SquareEmpty}}}
\multiput(0,1)(1,1){2}{\usebox{\SquareEmpty}} %
\put(1.5,1.5){\circle*{0.5}} \put(1.5,2.5){\circle{0.5}}
\end{picture}
&
\begin{picture}(4,4)(-0.5,-0.5) \DefineSquareEmpty
\multiput(1,0)(1,0){2}{\multiput(0,0)(0,1){2}{\usebox{\SquareEmpty}}}
\multiput(0,0)(1,2){2}{\usebox{\SquareEmpty}} %
\put(1.5,1.5){\circle*{0.5}} \put(1.5,2.5){\circle{0.5}}
\end{picture}
&
\begin{picture}(4,4)(-0.5,-0.5) \DefineSquareEmpty
\multiput(0,0)(1,0){2}{\multiput(0,0)(0,1){2}{\usebox{\SquareEmpty}}}
\multiput(2,0)(-2,2){2}{\usebox{\SquareEmpty}} %
\put(1.5,1.5){\circle*{0.5}} \put(1.5,0.5){\circle{0.5}}
\end{picture}
&
\begin{picture}(4,4)(-0.5,-0.5) \DefineSquareEmpty
\multiput(1,0)(1,0){2}{\multiput(0,0)(0,2){2}{\usebox{\SquareEmpty}}}
\multiput(0,1)(1,0){2}{\usebox{\SquareEmpty}} %
\put(1.5,1.5){\circle*{0.5}} \put(1.5,2.5){\circle{0.5}}
\end{picture}
\\   
\begin{picture}(4,4)(-0.5,-0.5) \DefineSquareEmpty
\multiput(1,0)(1,0){2}{\multiput(0,0)(0,2){2}{\usebox{\SquareEmpty}}}
\multiput(0,0)(1,1){2}{\usebox{\SquareEmpty}} %
\put(1.5,1.5){\circle*{0.5}} \put(1.5,2.5){\circle{0.5}}
\end{picture}
&
\begin{picture}(4,4)(-0.5,-0.5) \DefineSquareEmpty
\multiput(1,0)(1,0){2}{\multiput(0,0)(0,1){3}{\usebox{\SquareEmpty}}}
\put(0,1){\usebox{\SquareEmpty}} %
\put(1.5,1.5){\circle*{0.5}} \put(1.5,2.5){\circle{0.5}}
\end{picture}
&
\begin{picture}(4,4)(-0.5,-0.5) \DefineSquareEmpty
\multiput(1,0)(1,0){2}{\multiput(0,0)(0,1){3}{\usebox{\SquareEmpty}}}
\put(0,0){\usebox{\SquareEmpty}} %
\put(1.5,1.5){\circle*{0.5}} \put(1.5,2.5){\circle{0.5}}
\end{picture}
&
\begin{picture}(4,4)(-0.5,-0.5) \DefineSquareEmpty
\multiput(0,0)(1,-1){2}{\multiput(0,1)(1,0){2}{\multiput(0,0)(0,1){2}{\usebox{\SquareEmpty}}}} %
\put(1.5,1.5){\circle*{0.5}} \put(2.5,1.5){\circle{0.5}}
\end{picture}
&
\begin{picture}(4,4)(-0.5,-0.5) \DefineSquareEmpty
\multiput(0,0)(1,0){2}{\multiput(0,0)(0,2){2}{\usebox{\SquareEmpty}}}
\multiput(1,0)(1,0){2}{\multiput(0,0)(0,1){2}{\usebox{\SquareEmpty}}} %
\put(1.5,1.5){\circle*{0.5}} \put(1.5,2.5){\circle{0.5}}
\end{picture}
&
\begin{picture}(4,4)(-0.5,-0.5) \DefineSquareEmpty
\multiput(0,0)(1,0){3}{\multiput(0,0)(0,2){2}{\usebox{\SquareEmpty}}}
\put(1,1){\usebox{\SquareEmpty}} %
\put(1.5,1.5){\circle*{0.5}} \put(1.5,2.5){\circle{0.5}}
\end{picture}
&
\begin{picture}(4,4)(-0.5,-0.5) \DefineSquareEmpty
\multiput(0,0)(1,0){3}{\multiput(0,0)(0,1){2}{\usebox{\SquareEmpty}}}
\multiput(0,2)(1,0){2}{\usebox{\SquareEmpty}} %
\put(1.5,1.5){\circle*{0.5}} \put(2.5,1.5){\circle{0.5}}
\end{picture}
&
\begin{picture}(4,4)(-0.5,-0.5) \DefineSquareEmpty
\multiput(0,0)(1,0){3}{\multiput(0,0)(0,1){2}{\usebox{\SquareEmpty}}}
\multiput(0,2)(2,0){2}{\usebox{\SquareEmpty}} %
\put(1.5,1.5){\circle*{0.5}} \put(2.5,1.5){\circle{0.5}}
\end{picture}
\end{tabular}
\caption[Rightmost in first time step]{When $t = 0$ the first movement of an
agent located in $(x,y)$ should be decided according to initial contamination
status of the neighbors of $(x,y)$, as appears in these charts~---~the agent's
initial location is marked with a filled circle while the destination is marked
with an empty one. All configurations which do
not appear in these charts can be obtained by using rotations. This definition is needed in order to initialize the traversal behavior of the agents
in the correct direction.}%
\label{figure_rightmost}
\end{center}
\end{figure}
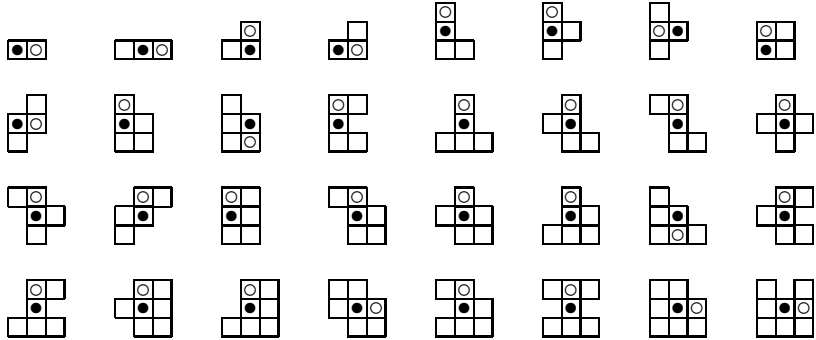

\begin{figure}[htbp]
\begin{center}
\epsfig{figure=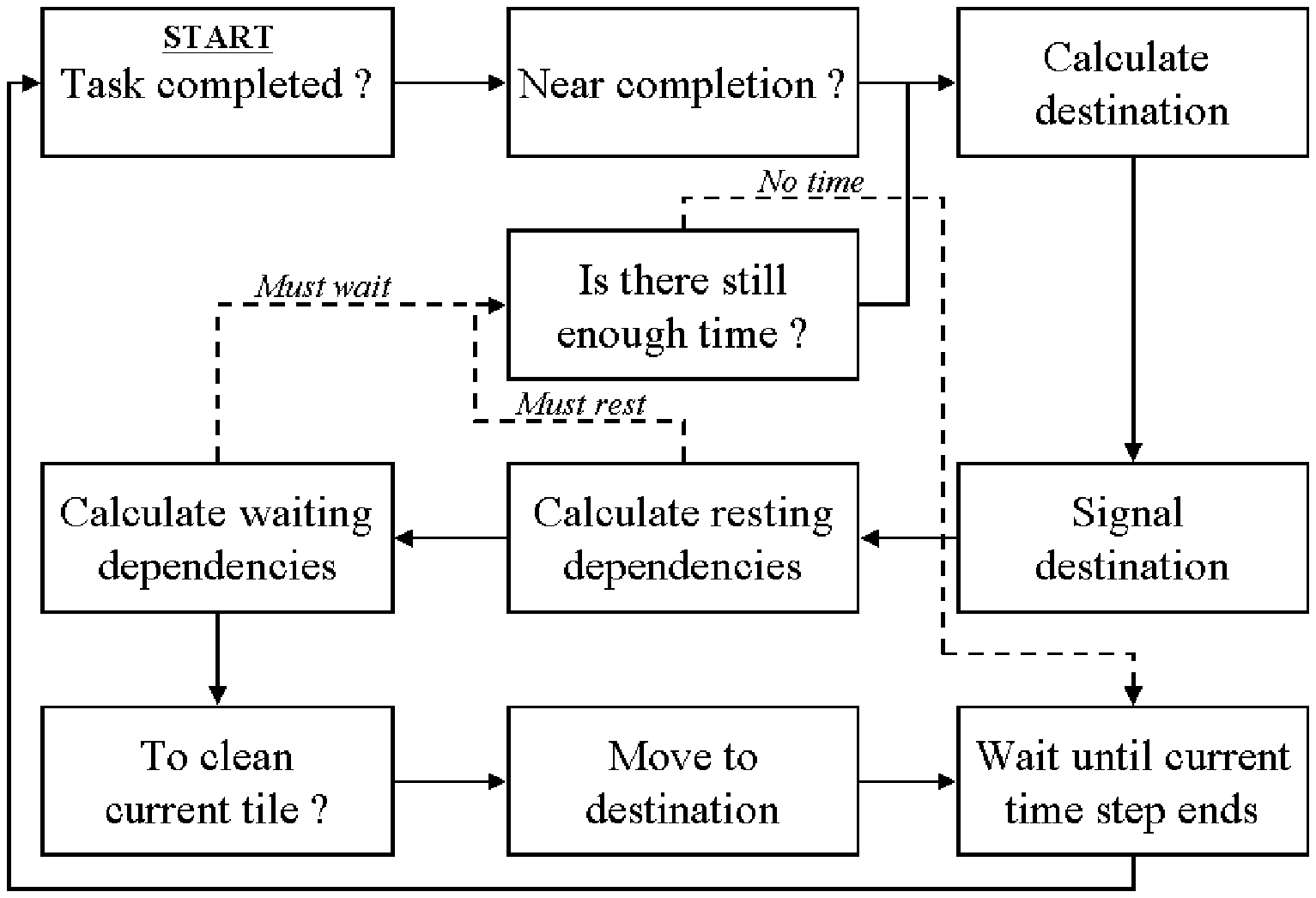,width=4in}
\end{center}
\caption[A schematic flow chart of the \textbf{SWEEP} protocol] %
{A schematic flow chart of the \textbf{SWEEP} protocol. The smooth lines
represent the basic flow of the protocol while the dashed lines represent cases
in which the flow is interrupted. Such interruptions occur when an agent
calculates that it must not move until other agents do so (either as a result
of \emph{waiting} or \emph{resting} dependencies~---~see
lines~\ref{clean.resting} and~\ref{clean.waiting_dependencies} of
\textbf{SWEEP} for more details).}%
\label{figure_SWEEP_FLOW}
\end{figure}

A schematic flowchart of the protocol, describing its major components and
procedures is presented in Figure~\ref{figure_SWEEP_FLOW}. The complete
pseudo-code of the protocol and its sub-routines appears in
Figures~\ref{figure_SWEEP1} and~\ref{figure_SWEEP2}. Upon initialization of the
system, the \emph{System Initialization} procedure is called (defined in
Figure~\ref{figure_SWEEP1}). This procedure sets various initial values of the
agents, and calls the protocol's main procedure~---~\emph{SWEEP} (defined in
Figure~\ref{figure_SWEEP2}). This procedure in turn, uses various sub-routines
and functions, all defined in Figure~\ref{figure_SWEEP1}.
The \emph{SWEEP} procedure is comprised of a loop which is executed continuously, until detecting one of two possible break conditions. The first, implemented in the \emph{Check Completion of Mission} procedure, is in charge of detecting cases where all the contaminated tiles have been cleaned. The second condition, implemented in the \emph{Check Near Completion of Mission} procedure, is in charge of detecting scenarios in which every contaminated tile contains at least a single agent. In this case, the next operation would be a simultaneous cleaning of the entire contaminated tiles. %
Until these conditions are met, each agent goes through the following sequence of commands. First each agent calculates its desired destination at the current turn. Then, each agent calculated whether it should give a priority to another agent located at the same tile, and wishes to move to the same destination. When two or more agents are located at the same tile, and wish to move towards the same direction, the agent who had entered the tile first gets to leave the tile, while the other agents wait. In case several agents had entered the tile at the same time, the priority is determined using the \emph{Priority} function. Before actually moving, each agent who had obtained a permission to move, must now locally synchronize its movement  with its neighbors, in order to avoid simultaneous movements which may damage the connectivity of the region. This is done using the \emph{waiting dependencies} mechanism, which is implemented by each agent via an internal positioning of itself in a local ordering of his neighboring agents. When an agent is not delayed by any other agent, it executes its desired movement.
It is important to notice that at any given time, \emph{waiting} or \emph{resting} agents may become active again, if the conditions which made them become inactive in the first place, had changed.

\begin{figure}
\begin{scriptsize}
\begin{algorithmic}[1]
\STATE \textbf{System Initialization} 
\STATE $\qquad$ Arbitrarily choose a \emph{pivot point} $p_{0}$ in $\partial F_{0}$, and mark it as \emph{critical point} %
\STATE $\qquad$ Place all the agents on $p_{0}$ %
\STATE $\qquad$ \textbf{For} ($i=1$; $i \leq k$; $i++$) do %
\STATE $\qquad$ $\qquad$ Call \textbf{Agent Reset} for agent i %
\STATE $\qquad$ $\qquad$ Call \textbf{SWEEP} for agent i %
\STATE $\qquad$ $\qquad$ Wait two time steps %
\STATE $\qquad$ \textbf{End for} %
\STATE \textbf{End procedure} \\ $\ $\\ %
\STATE \textbf{Agent Reset} 
\STATE $\qquad$ \emph{resting} $\leftarrow$ \emph{false} %
\STATE $\qquad$ \emph{dest} $\leftarrow$ null  /* destination */ %
\STATE $\qquad$ \emph{near completion} $\leftarrow$ \emph{false} %
\STATE $\qquad$ \emph{saturated perimeter} $\leftarrow$ \emph{false} %
\STATE $\qquad$ \emph{waiting} $\leftarrow$ $\emptyset$ %
\STATE \textbf{End procedure} \\ $\ $\\ %
\STATE \textbf{Priority} %
\STATE $\qquad$ /* Assuming the agent moved from $(x_{0},y_{0})$ to $(x_{1},y_{1})$ */
\STATE $\qquad$ \emph{priority} $\leftarrow$ $2(x_{1}-x_{0}) + (y_{1}-y_{0})$ %
\STATE \textbf{End procedure} \\ $\ $\\ %
\STATE \textbf{Check Completion of Mission} %
\STATE $\qquad$\textbf{If} ($(x,y) = p_{0}$) and $(x,y)$ has no contaminated neighbors then %
\STATE $\qquad \qquad$ \textbf{If} $(x,y)$ is contaminated then %
\STATE $\qquad \qquad \qquad$ Clean $(x,y)$ %
\STATE $\qquad \qquad$ \textbf{STOP} %
\STATE \textbf{End procedure} \\ $\ $\\ %
\STATE \textbf{Check ``Near Completion'' of Mission} %
\STATE $\qquad$ /* Cases where every tile in $F_{t}$ contains at least a single agent */ %
\STATE $\qquad$ \emph{near completion} $\leftarrow$ \emph{false} %
\STATE $\qquad$ \textbf{If} each of the contaminated neighbors of (x,y) contains at least one agent then %
\STATE $\qquad \qquad$  \emph{near completion} $\leftarrow$ \emph{true}%
\STATE $\qquad$ \textbf{If} each of the contaminated neighbors of (x,y) satisfies \emph{near completion} then  %
\STATE $\qquad \qquad$ Clean $\left(x,y\right)$ and \textbf{STOP}%
\STATE $\qquad$ /* Cases where every non-critical tile in $\partial F_{t}$ contains at least 2 agents */ %
\STATE $\qquad$ \emph{saturated perimeter} $\leftarrow$ \emph{false} %
\STATE $\qquad$ \textbf{If} $\left(\left(x,y\right) \in \partial F_{t}\right)$ and both $\left(x,y\right)$ and all of its non-critical neighbors \\ $\qquad$ in $\partial F_{t}$ contain at least two agents then %
\STATE $\qquad \qquad$  \emph{saturated perimeter} $\leftarrow$ \emph{true} %
\STATE $\qquad$ \textbf{If} $\left(\left(x,y\right) \in \partial F_{t}\right)$ and both $\left(x,y\right)$ and all of its neighbors in $\partial F_{t}$ has \\ $\qquad$ \emph{saturated perimeter = true} then %
\STATE $\qquad \qquad$ Ignore \emph{resting} commands for this time step %
\STATE \textbf{End procedure} %
\end{algorithmic}
\end{scriptsize}
\caption[The first part of the \textbf{SWEEP} cleaning protocol] %
{The first part of the \textbf{SWEEP} cleaning protocol.}%
\label{figure_SWEEP1}
\end{figure}

\begin{figure}
\begin{scriptsize}
\begin{algorithmic}[1]
\STATE \textbf{SWEEP Protocol} /* Controls agent $i$ after \textbf{Agent Reset}  */ %
\STATE $\quad$ \textbf{Check Completion of Mission} \label{clean.1} %
\STATE $\quad$ \textbf{Check ``Near Completion'' of Mission} %
\STATE $\quad$ \emph{dest} $\leftarrow$ \emph{rightmost neighbor} of (x,y) /* Calculate destination */ \label{clean.destination} %
\STATE $\quad$ \emph{destination signal bits} $\leftarrow$ \emph{dest} /* Signaling the desired destination */ \label{clean.signaling} %
\STATE $\quad$ /* Calculate resting dependencies (solves agents' clustering problem) */ \label{clean.resting} %
\STATE $\quad$ From all agents in (x,y) except agent $i$ we define be the following groups: %
\STATE $\quad$ $\quad$ $K_{1}$ : Agents signaling towards \emph{dest} which entered (x,y) before agent $i$ %
\STATE $\quad$ $\quad$ $K_{2}$ : Agents signaling towards \emph{dest} which entered (x,y) with agent $i$, \\ $\quad$ $\quad$ $\qquad$ and with higher \emph{priority} than this of agent $i$ %
\STATE $\quad$ \emph{resting} $\leftarrow$ \emph{false} %
\STATE $\quad$ \textbf{If} ($K_{1} \neq \emptyset$) or ($K_{2} \neq \emptyset$) then %
\STATE $\quad$ $\quad$ \emph{resting} $\leftarrow$ \emph{true} %
\STATE $\quad$ $\quad$ \textbf{If} (current time-step $\mathcal{T}$ did not end yet) then jump to \ref{clean.destination} \textbf{Else} jump to \ref{clean.last} \label{clean.resting2} %
\STATE $\quad$ \emph{waiting} $\leftarrow$ $\emptyset$ /* Waiting dependencies (agents synchronization) */ \label{clean.waiting_dependencies} %
\STATE $\quad$ Let \emph{active agent} denote a \emph{non-resting} agent which didn't move in $\mathcal{T}$ yet %
\STATE $\quad$ \textbf{If} (x-1,y) $ \in F_{t}$ contains an active agent then \emph{waiting} $\leftarrow$ \emph{waiting} $\cup$ \{\emph{left}\} %
\STATE $\quad$ \textbf{If} (x,y-1) $ \in F_{t}$ contains an active agent then \emph{waiting} $\leftarrow$ \emph{waiting} $\cup$ \{\emph{down}\} %
\STATE $\quad$ \textbf{If} (x-1,y-1) $\in F_{t}$ contains an active agent then \emph{waiting} $\leftarrow$ \emph{waiting} $\cup$ \{\emph{l-d}\} %
\STATE $\quad$ \textbf{If} (x+1,y-1) $\in F_{t}$ contains an active agent then \emph{waiting} $\leftarrow$ \emph{waiting} $\cup$ \{\emph{r-d}\} %
\STATE $\quad$ \textbf{If} \emph{dest} = \emph{right} and (x+1,y) contains an active agent $j$, and \emph{dest}$_{j} \neq$ \emph{left}, and \\ $\quad$ there are no other agents delayed by agent $i$ (i.e. (x-1,y) does not contain \\ $\quad$  active agent $l$ with \emph{dest}$_{l} = $\emph{right} and no active agents in (x,y+1),(x+1,y+1), \\ $\quad$ (x-1,y+1), and (x+1,y) does not contain active agent $n$ with \emph{dest}$_{n} =$ \emph{left}), \\ $\quad$ then (\emph{waiting} $\leftarrow$ \emph{waiting} $\cup$ \{\emph{right}\}) and $\left(\emph{waiting}_{j} \leftarrow \emph{waiting}_{j} \setminus \{\emph{left}\}\right)$ %
\STATE $\quad$ \textbf{If} \emph{dest} = \emph{up} and (x,y+1) contains an active agent $j$, and \emph{dest}$_{j} \neq$ \emph{down}, and \\ $\quad$ there are no other agents delayed by agent $i$ (i.e. (x,y-1) does not contain \\ $\quad$ active agent $l$ with \emph{dest}$_{l} = $\emph{up} and no active agents in (x+1,y),(x+1,y+1), \\ $\quad$  (x-1,y+1), and (x,y+1) does not contain active agent $n$ with \emph{dest}$_{n} =$ \emph{down}), \\ $\quad$ then (\emph{waiting} $\leftarrow$ \emph{waiting} $\cup$ \{\emph{up}\}) and $\left(\emph{waiting}_{j} \leftarrow \emph{waiting}_{j} \setminus \{\emph{down}\}\right)$   \label{clean.waiting_dependencies2} %
\STATE $\quad$ \textbf{If} (\emph{waiting} $\neq \emptyset$) then %
\STATE $\quad$ $\quad$ \textbf{If} ($\mathcal{T}$ has not ended yet) then jump to \ref{clean.destination} \textbf{Else} jump to \ref{clean.last} %
\STATE $\quad$ /* Decide whether or not (x,y) should be cleaned */  %
\STATE $\quad$ \textbf{If} $\neg$ ((x,y) $\in \partial F_{t}$) or ((x,y) $\equiv p_{0}$) or (x,y) has 2 contaminated tiles in its \\
$\quad$ $\FourNeigh{}$ which are not connected via a path of contaminated tiles from its \\
$\quad$ $\EightNeigh$ then %
\STATE $\quad$ $\quad$ (x,y) is an \emph{internal point} or a \emph{critical point} and should not be cleaned %
\STATE $\quad$ \textbf{Else} %
\STATE $\quad$ $\quad$ Clean (x,y) if and only if it does not still contain other agents \label{clean.not_last_agent} %
\STATE $\quad$ Move to \emph{dest} %
\STATE $\quad$ Wait until $\mathcal{T}$ ends.\label{clean.last} %
\STATE $\quad$ Return to \ref{clean.1} %
\end{algorithmic}
\end{scriptsize}
\caption[The \textbf{SWEEP} cleaning protocol] %
{The \textbf{SWEEP} cleaning protocol.}%
\label{figure_SWEEP2}
\end{figure}

Following are several results that we will later use. While using these results, we note that completely \emph{cleaning}
a region is at least as strong as \emph{covering} it, as the number of ``uncovered'' tiles at
any given time can be modeled by the number of ``contaminated'' tiles, since the
number of uncovered tiles in the original region to be explored is clearly upper bounded at all times by the number of
remaining contaminated tiles that belong to this region.

\begin{result} \textbf{(Cleaning a Non-Expanding Contamination)}
\label{theorem1}
The time it takes for a group of $K$ robots using the \textbf{SWEEP} algorithm to clean a region $F$ of the grid is at most:
\[
t_{static} \triangleq \frac{8(| \partial F_{0} | - 1) \cdot (W(F_{0}) + k)}{k} + 2k
\]
\end{result}
Here $W(F)$ denotes the depth of the region $F$ (the shortest path from some internal point in $F$ to its boundary, for the internal point whose shortest path is the longest) and as defined above, $\partial F$ denotes the boundary of $F$, defined via~:
\begin{displaymath}
\partial F = \{v \ | \ v \in F \ \wedge \ 8-Neighbors(v) \ \cap \
(G \ \setminus \ F) \ \neq \ \emptyset\}
\end{displaymath}

The term $8-Neighbors(v)$ is used to denote the eight tiles that tile $v$ is immediately surrounded by.

\begin{result} \textbf{(Universal Lower Bound on Contaminated Area)}
\label{theorem3}
Using any cleaning algorithm, the area at time $t$ of a contaminated region that expands every $d$ time steps can be recursively lower bounded, as follows~:
\begin{displaymath}
S_{t+d} \geq S_{t} - d \cdot k + \left\lfloor 2\sqrt{2 \cdot (S_{t} - d \cdot k) - 1} \right\rfloor
\end{displaymath}
\end{result}
Here $S_{t}$ denotes the area of the contaminated region at time $t$ (such that $S_{0} = n$).

\begin{result} \textbf{(Upper Bound on Cleaning Time for SWEEP on Expanding Domains)}
\label{theorem2}
For a group of $k$ robot using the \textbf{SWEEP} algorithm to clean a region $F$ on the grid, that expands every $d$ time steps, the time it takes the robots to clean $F$ is at most $d$ multiplied by the minimal positive value of the following two numbers~:
\begin{displaymath}
\frac{(A_{4} - A_{1} A_{3}) \pm \sqrt{(A_{1} A_{3} -
A_{4})^{2} - 4 A_{3} (A_{2} - A_{1} - A_{1} A_{4})}}{2 A_{3}}
\end{displaymath}
where~:
\begin{displaymath}
A_{1} = \frac{c_{0} + 2 - \gamma_{2}}{4}
\ , \
A_{2} = \frac{c_{0} + 2 + \gamma_{2}}{4}
\ , \
A_{3} = \frac{8 \cdot \gamma_{2}}{d \cdot k}
\ , \
\end{displaymath}
\begin{displaymath}
A_{4} = \gamma_{1} - \frac{\gamma_{2} \cdot \gamma}{d}
\ \ , \ \
\gamma_{1} = \psi \left(1 + A_{2} \right) - \psi
\left(1 + A_{1} \right)
\
\end{displaymath}
\begin{displaymath}
\gamma_{2} = \sqrt{(c_{0} + 2)^{2} - 8 S_{0} + 8}
\end{displaymath}
\begin{displaymath}
\gamma = \frac{8(k + W(F_{0}))}{k} - \frac{d - 2k}{|\partial F_{0}| -
1}
\end{displaymath}
\end{result}

Here $c_{0}$ is the circumference of the initial region $F_{0}$, and where $\psi(x)$ is the \emph{Digamma} function (studied in~\cite{Digamma})~---~the
logarithmic derivative of the \emph{Gamma function}, defined as~:
\[
\psi(x)=\frac{d}{dx} \ln \Gamma(x) = \frac{\Gamma'(x)}{\Gamma(x)}
\]
or as~:
\[
\psi(x)= \int_0^{\infty}\left(\frac{e^{-t}}{t} - \frac{e^{-xt}}{1 -
e^{-t}}\right)\,dt
\]

Note that although $c_{0} = O(|\partial F|)$ the actual length of the perimeter of the region can be greater than its cardinality, as several tiles may be traversed more than once. In fact, in \cite{CC08} it is shown that $c_{0} \leq 2 \cdot |\partial F| - 2$.

\section{Grid Coverage --- Analysis}

We note again that when discussing the coverage of regions on the grid, either static or expanding, it is enough to show that the region can be cleaned by the team of robots, as clearly the cleaned sites are always a subset of the visited ones.
We first present the cover time of a group of robots operating in non-expanding domains, using the \textbf{SWEEP} algorithm.

\begin{theorem}
\label{thm1}
Given a connected region of $S_{0} = n$ grid tiles and perimeter $c_{0}$, that should be covered by a team of $k$ ant-like robots, the robots can cover it using $O \left( \frac{1}{k} S_{0}^{1.5} + S_{0} \right)$ time.
\begin{proof}
Since $| \partial F_{0} | = \Theta(c_{0})$, and $W(F_{0}) = O(\sqrt{S_{0}})$, recalling Result~\ref{theorem1} we can see that~:
\[
t_{k}(n) = t_{static}(k) = O \left( \frac{1}{k} \sqrt{S_{0}} \cdot c_{0} + c_{0} + k \right)
\]

As $c_{0} = O(S_{0})$ and as for practical reasons we assume that $k < n$ this would equal in the worse case to~:
\[
t_{k}(n) = t_{static}(k) = O \left( \frac{1}{k} S_{0}^{1.5} + S_{0} \right)
\]
\end{proof}
\end{theorem}

We now examine the problem of covering expanding domains. The lower bound for the number of robots required for completing is as follows.

\begin{theorem}
\label{thm2}
Given a region of size $S_{0} \geq 3$ tiles, expanding every $d$ time steps, then a team of less than $\frac{\sqrt{S_{0}}}{d}$ robots cannot clean the region, regardless of the algorithm used.
\begin{proof}
Recalling Result~\ref{theorem3} we can see that~:
\[
S_{t+d} - S_{t} \geq  \left\lfloor 2\sqrt{2 \cdot (S_{t} - d \cdot k) - 1} \right\rfloor - d \cdot k
\]

By assigning $k = \frac{\sqrt{S_{0}}}{d}$ we can see that~:
\[
\Delta S_{t} = S_{t+d} - S_{t} \geq  \left\lfloor 2\sqrt{2 \cdot (S_{t} - \sqrt{S_{0}}) - 1} \right\rfloor - \sqrt{S_{0}}
\]

For any $S_{0} \geq 3$, we see that $\Delta S_{0} > 0$. In addition, for every $S_{0} \geq 3$ we can see that $\frac{d S_{t}}{d t} > 0$ for every $t \geq 0$. Therefore, for every $S_{0} \geq 3$ the size of the region will be forever growing.
\end{proof}
\end{theorem}

\begin{corollary}
\label{thm2.cor}
Given a region of size $S_{0}$ tiles, expanding every $d$ time steps, where $d = O(1)$ w.r.t $S_{0}$, then a team of less than $\Omega(\sqrt{S_{0}})$ robots cannot clean the region, regardless of the algorithm used.
\end{corollary}

\begin{theorem}
\label{thm.new.lower}
Given a region $F$ of size $S_{0}$ tiles, expanding every $d$ time steps, where $R(F)$ is the perimeter of the bounding rectangle of the region $F$, then a team of $k$ robots that at $t = 0$ are located at the same tile cannot clean the region, regardless of the algorithm used, as long as $d^{2} k < \Omega(R(F))$.
\begin{proof}
For every $v \in F$ let $l(v)$ denote the maximal distance between $v$ and any of the tiles of $F$, namely~:
\[
l(v) = \max\{ d(v,u) | u \in F\}
\]
Let $C(F) = l(v_{c})$ such that $v_{c} \in F$ is the tile with minimal value of $l(v)$.

Let $v_{s}$ denote the tile the agents are located in at $t = 0$. Let $v_{d} \in F$ denote some contaminated tile such that $d(v_{s},v_{d}) = l(v_{s})$. Regardless of the algorithm used by the agents, until some agent reaches $v_{d}$ there will pass at least $l(v_{s})$ time steps. Let us assume w.l.o.g that $v_{d}$ is located to the right (or of the same horizontal coordinate) and to the top (or of the same vertical coordinate) of $v_{s}$. Then by the time some agent is able to reach $v_{d}$ there exists an upper-right quarter of a digital sphere of radius $\left\lfloor\frac{l(v_{s})}{d}\right\rfloor + 1$, whose center is $v_{d}$.

The number of tiles in such a quarter of digital sphere equals~:
\[
\frac{1}{2} \left\lfloor\frac{l(v_{s})}{d}\right\rfloor^{2} + \frac{3}{2} \left\lfloor\frac{l(v_{s})}{d}\right\rfloor + 1 = \Theta\left(\frac{l(v_{s})^{2}}{d^{2}}\right)
\]

It is obvious that the region cannot be cleaned until $v_{d}$ is cleaned. Let $t_{d}$ denote the time at which the first agent reaches $v_{d}$. It is easy to see that $t_{d} \geq l(v_{s})$. Therefore, regardless of activities of the agents until time step $t_{d}$, there are now $k$ agents that has to clean a region of at least $\Theta\left(\frac{l(v_{s})^{2}}{d^{2}}\right)$ tiles, spreading every $d$ time steps. Using Theorem~\ref{thm2} we know that $k$ agents cannot clean an expanding region of $k = \frac{\sqrt{S_{0}}}{d}$ tiles. Namely, at time $t_{d}$ $k$ agents could not clean the contaminated tiles if~:
\[
d^{2} k < \Omega\left(l(v_{s})\right)
\]

As $l(v_{s}) \geq C(F)$ we know that $k$ agents could not clean an expanding contaminated region where~:
$d^{2} k < \Omega\left(C(F)\right)$. It is easy to see that for every region $F$, if $R(F)$ is the length of the perimeter of the bounding rectangle of $F$ then $C(F) = \Theta(R(F))$.
\end{proof}
\end{theorem}

\begin{lemma}
\label{lemma.gamma2}
For every connected region of size $S_{0} \geq 3$ and perimeter of length $c_{0}$~:
\[
\frac{1}{2} c_{0} < \gamma_{2} < c_{0}
\]
\begin{proof}
let us assume by contradiction that $(c_{0} + 2)^{2} \leq (8 S_{0} + 8)$. This means $c_{0} \leq \sqrt{8 S_{0} + 8} - 2$. However, the minimal circumference of a region of size $S_{0}$ is achieved when the region is arranged in the form of an 8-connected digital sphere, in which case $c_{0} \geq 4 \sqrt{S_{0}} - 4$, contradicting the assumption that $c_{0} \leq \sqrt{8 S_{0} + 8} - 2$ for every $S_{0} > 5$. Therefore, $\gamma_{2} \in \mathbb{R}$.

Let us assume by contradiction that $\gamma_{2} < \frac{1}{2} c_{0}$. Therefore~:
\[
(c_{0} + 2)^{2} - 8 S_{0} + 8 < \frac{1}{4} c_{0}^{2}
\]
which implies~:
\[
c_{0} < -\frac{16}{6} + \sqrt{10\frac{2}{3} S_{0} - 8\frac{8}{9}} < 3.266 \sqrt{S_{0}} - 2
\]
However, we know that $c_{0} \geq 4 \sqrt{S_{0}} - 4$, which contradicts the assumption that $\gamma_{2} < \frac{1}{2} c_{0}$ for every $S_{0} \geq 3$.

Let us assume by contradiction that $\gamma_{2} > c_{0}$. Therefore~:
\[
(c_{0} + 2)^{2} - 8 S_{0} + 8 > c_{0}^{2}
\]
which implies~:
\[
c_{0} >  4 S_{0} - 6
\]
However, we know that $c_{0} \leq 2 S_{0} - 2$ (as $c_{0}$ is maximized when the tiles are arranged in the form of a straight line), contradicting the assumption that $\gamma_{2} >  c_{0}$ for every $S_{0} \geq 3$.
\end{proof}
\end{lemma}

\begin{lemma}
\label{lemma.gamma1}
For every connected region of size $S_{0} \geq 3$ and perimeter of length $c_{0}$~:
\[
\Omega(1) < \gamma_{1} < O(\ln n)
\]
\begin{proof}
Let us observe $\gamma_{1}$~:
\begin{displaymath}
\gamma_{1} \triangleq \psi \left(1 + \frac{c_{0} + 2 + \gamma_{2}}{4} \right) -
\psi \left(1 + \frac{c_{0} + 2 - \gamma_{2}}{4} \right)
\end{displaymath}
From Lemma \ref{lemma.gamma2} we can see that $1 < \left(1 + \frac{c_{0} + 2 - \gamma_{2}}{4} \right) < \frac{1}{4} c_{0}$. Note that
$\psi(1) = -\hat{\gamma}$ where $\hat{\gamma}$ is the \emph{Euler–-Mascheroni}
constant, defined as~:
\[
\hat{\gamma} = \lim_{n \rightarrow \infty } \left[ \left( \sum_{k=1}^n
\frac{1}{k} \right) - \log(n) \right]=\int_1^\infty\left({1\over\lfloor
x\rfloor}-{1\over x}\right)\,dx
\]
which equals approximately $0.57721$. In addition, $\psi(x)$ is monotonically
increasing for every $x > 0$. As we also know that $\psi(x)$ is upper bounded
by $O(\ln x)$ for large values of $x$, we see that~:
\begin{equation}\label{eq.complexityDynamicPsi1}
-0.58 < \psi \left(1 + \frac{c_{0} + 2 - \gamma_{2}}{4} \right) < O(\ln n)
\end{equation}

From Lemma \ref{lemma.gamma2} we also see that $1 < \left(1 + \frac{c_{0} + 2 + \gamma_{2}}{4} \right) < \frac{1.5}{4} c_{0}$ meaning that~:
\begin{equation}\label{eq.complexityDynamicPsi2}
\psi \left(1 + \frac{c_{0} + 2 + \gamma_{2}}{4} \right) = \Theta(\ln n)
\end{equation}

Combining equations \ref{eq.complexityDynamicPsi1} and
\ref{eq.complexityDynamicPsi2} we see that~:
\begin{equation}\label{eq.complexityDynamicPsi3}
\Omega(1) < \gamma_{1} < O(\ln n)
\end{equation}
\end{proof}
\end{lemma}

\begin{theorem}
\label{theorem.thereisanswer}
Result~\ref{theorem2} returns a positive real number for the covering time of a region of $S_{0}$ tiles that expands every $d$ time steps, when the number of robots is $\Theta(\sqrt{S_{0}})$ and $d = \Omega(\frac{c_{0}}{\gamma_{1}})$, and where $\gamma_{1}$ shifts from $O(1)$ to $O(\ln S_{0})$ as $c_{0}$ grows from $O(\sqrt{S_{0}})$ to $O(S_{0})$, defined as~:
\[
\gamma_{1} = \psi \left(1 + \frac{c_{0} + 2 + \gamma_{2}}{4} \right) - \psi
\left(1 + \frac{c_{0} + 2 - \gamma_{2}}{4} \right)
\]
\[
\gamma_{2} = \sqrt{(c_{0} + 2)^{2} - 8 S_{0} + 8}
\]
\begin{proof}
Following are the requirements that must hold in order for Result~\ref{theorem2} to yield a real number~:
\begin{itemize}
  \item $d \cdot k \neq 0$
  \item $| \partial F | > 1$
  \item $A_{3} \neq 0$
  \item $(c_{0} + 2)^{2} > 8 S_{0} - 8$
  \item $(A_{1} A_{3} - A_{4})^{2} \geq 4 A_{3} (A_{2} - A_{1} - A_{1} A_{4})$
\end{itemize}

The first and second requirements hold for every non trivial scenario. The third requirement is implied by the fourth. The fourth assumption is a direct result of Lemma~\ref{lemma.gamma2}.

As for the last requirement, we ask that~:
\[A_{1}^{2} A_{3}^{2} + A_{4}^{2} \geq 4 A_{2} A_{3} - 4 A_{1} A_{3} - 2 A_{1} A_{3} A_{4}\]
which subsequently means that we must have~:
\[
\begin{array}{c}
\frac{\gamma_{2}^{2}}{d^{2} k^{2}} \left( c_{0}^{2} + \gamma_{2}^{2} - c_{0} \gamma_{2} \right) \\ + \gamma_{1}^{2} + \frac{\gamma_{2}^{2} \gamma^{2}}{d} - \frac{\gamma \gamma_{1} \gamma_{2}}{d}
\end{array}
\geq 4 \frac{\gamma_{2}}{d k}
\left(\begin{array}{c}
4 \gamma_{2} - c_{0} \gamma_{1} + \\ c_{0} \frac{\gamma_{2} \gamma}{d} - 2 \gamma_{1} +  2 \frac{\gamma_{2} \gamma}{d} + \\ \gamma_{1} \gamma_{2} - \frac{\gamma_{2}^{2} \gamma}{d}
\end{array}
\right)\]

Using Lemma \ref{lemma.gamma2} and Lemma \ref{lemma.gamma1} we should make sure that~:

\[
\frac{\gamma_{2}^{2}}{d k^{2}} c_{0}^{2}  +  \gamma_{1}^{2} d + \gamma_{2}^{2} \gamma^{2} - \gamma \gamma_{1} \gamma_{2}
\geq
O \left( \frac{c_{0} \gamma_{2} \gamma_{1}}{k} + \frac{c_{0} \gamma_{2}^{2} \gamma}{dk} \right)
\]

Using $W(F) = O(\sqrt{S_{0}})$ and $\Omega(\sqrt{S_{0}}) = | \partial F | = O(S_{0})$ and dividing by $\gamma_{2}^{2}$ (which we know to be larger than zero),  we can write the above as follows~:
\[
\frac{c_{0}^{2}}{d k^{2}}  +  \frac{k^{2} + d \ln^{2} S_{0}}{c_{0}^{2}} + 1
\geq
\]
\[O\left(
\frac{\ln S_{0}}{c_{0}} + \frac{k \ln S_{0}}{c_{0}^{2}} + \frac{\ln S_{0}}{k} + \frac{c_{0} \sqrt{S_{0}}}{d k^{2}} + \frac{c_{0} }{d k} + \frac{1}{d}
\right)
\]


As $c_{0} \geq \sqrt{S_{0}}$ then $\frac{c_{0}^{2}}{d k^{2}} \geq \frac{c_{0} \sqrt{S_{0}}}{d k^{2}}$. In addition, $1 \geq \frac{1}{d}$ and also $1 \geq \frac{\ln S_{0}}{c_{0}}$ and $1 \geq \frac{\ln S_{0}}{k}$ (as Eq. \ref{eq.theorem.agents3} shows that $k \geq \ln S_{0}$). In order to have also $1 \geq \frac{c_{0}}{d k}$ we must have~:

\begin{equation}
\label{eq.theorem.agents1}
d \cdot k = \Omega(c_{0})
\end{equation}

In addition, we should also require that the result $\mu$ would be positive (as it denotes the coverage time). Namely, that~:
\[A_{4} + \sqrt{(A_{1} A_{3} - A_{4})^{2} - 4 A_{3} (A_{2} - A_{1} - A_{1} A_{4})} > A_{1}A_{3}\]

For this to hold we shall merely require that:
\[
A_{2} - A_{1} - A_{1} A_{4} \leq 0
\]
(as $A_{3}$ is known to be positive).
Assigning the values of $A_{1}, A_{2}, A_{4}$, this translates to~:

\[
c_{0} + c_{0}^{2} \frac{\gamma}{d} \leq O(c_{0} \gamma_{1})
\]

Dividing by $c_{0}$ we can write~:
\[
1 + c_{0} \frac{1 + \frac{\sqrt{S_{0}}}{k} - \frac{d}{c_{0}} + \frac{k}{c_{0}}}{d} \leq  O(\gamma_{1})
\]

Namely~:
\[
c_{0} + \frac{c_{0} \sqrt{S_{0}}}{k} + k \leq  d O(\gamma_{1})
\]

As $c_{0}$ is the dominant element of the left side of the inequation, we see that~:
\begin{equation}
\label{eq.theorem.agents2}
d = \Omega\left(\frac{c_{0}}{\gamma_{1}}\right)
\end{equation}

Assigning this lower bound for $d$ we can now see that~:
\begin{equation}
\label{eq.theorem.agents3}
\Omega(\sqrt{S_{0}}) \leq k \leq O(c_{0})
\end{equation}

Therefore, we shall select the value of $k$ such that~:
\[
k = \Theta(\sqrt{S_{0}})
\]

This also satisfies Equation \ref{eq.theorem.agents1}.
\end{proof}
\end{theorem}

\begin{theorem}
\label{theorem.time}
The time it takes a group of $k = \Theta(\sqrt{S_{0}})$ robots using the \textbf{SWEEP} algorithm to cover a connected region of size $S_{0}$ tiles, that expands every $d = \Omega(\frac{c_{0}}{\gamma_{1}})$ time steps, is upper bounded as follows~:
\[t_{SUCCESS} =
O \left( S_{0}^{2} \ln S_{0} \right)
\]

where $\gamma_{1}$ shifts from $O(1)$ to $O(\ln S_{0})$ as $c_{0}$ grows from $O(\sqrt{S_{0}})$ to $O(S_{0})$, defined as~:
\[
\gamma_{1} = \psi \left(1 + \frac{c_{0} + 2 + \gamma_{2}}{4} \right) - \psi
\left(1 + \frac{c_{0} + 2 - \gamma_{2}}{4} \right)
\]
\[
\gamma_{2} = \sqrt{(c_{0} + 2)^{2} - 8 S_{0} + 8}
\]

\begin{proof}
Recalling Result~\ref{theorem1} we know that if~:
\[
\frac{8(| \partial F_{0} | - 1) \cdot (W(F_{0}) + k)}{k} + 2k < d
\]
then the robots could clean the region before it expands even once. In this case, the cleaning time would be $O(\frac{1}{k} \sqrt{S_{0}} \cdot c_{0} + c_{0})$ as was shown in Theorem~\ref{thm1}. Therefore, we shall assume that~:
\begin{equation}\label{eq.d_k_connection1}
\frac{8(| \partial F_{0} | - 1) \cdot (W(F_{0}) + k)}{k} + 2k \geq d
\end{equation}

Observing Result~\ref{theorem2} we see that~: \\
$t_{SUCCESS} =$
\[
d \cdot O \left(A_{1} + \frac{|A_{4}|}{A_{3}} + \sqrt{A_{1}^{2} +
\frac{|A_{1}A_{4}| + A_{1} + A_{2}}{A_{3}} + \frac{A_{4}^{2}}{A_{3}^{2}}} \right)
\leq
\]
\[
d \cdot O \left(A_{1} + \frac{|A_{4}|}{A_{3}} +
\frac{\sqrt{A_{3}}\sqrt{|A_{1}A_{4}| + A_{1} + A_{2}}}{A_{3}} +
\frac{|A_{4}|}{A_{3}} \right) \leq
\]
\[
d \cdot O \left(A_{1} + \frac{|A_{4}|}{A_{3}} + \frac{\sqrt{|A_{1}A_{4}|} +
\sqrt{A_{1}} + \sqrt{A_{2}} }{\sqrt{A_{3}}} \right) =
\]
%
%
\[
d \cdot O \left(    \begin{array}{c}
                      c_{0} + \gamma_{2} + dk\frac{\gamma_{1}}{\gamma_{2}} + k\gamma +  \\
                      \sqrt{k}\frac{\sqrt{c_{0} + \gamma_{2}}\sqrt{d\gamma_{1}
                        + \gamma_{2} \cdot \gamma}}{\sqrt{\gamma_{2}}} + \sqrt{\frac{k d}{\gamma_{2}}} \sqrt{c_{0} + \gamma_{2}}
                    \end{array}
\right)
\]

Using the fact that $\gamma_{2} = \Theta(c_{0})$ (Lemma~\ref{lemma.gamma2}) we can rewrite this expression as~:
\begin{equation}\label{eq.complexityDynamic1}
d \cdot O \left(c_{0} + dk \frac{\gamma_{1}}{c_{0}} + k\gamma
+ \sqrt{k} \sqrt{ d\gamma_{1} + c_{0} \gamma} + \sqrt{k d}
\right)
\end{equation}

Recalling Equation \ref{eq.d_k_connection1}, and as $W(F_{0}) = O(\sqrt{S_{0}})$, we can see that~:
\[
d = O \left(\frac{\sqrt{S_{0}} \cdot c_{0}}{k} + c_{0} + k \right)
\]

Therefore, $|\gamma|$ can now be written as~:
\begin{displaymath}
|\gamma| = O\left(\frac{\sqrt{S_{0}}}{k} + \sqrt{S_{0}} + \frac{k}{\sqrt{S_{0}}} + 1\right)
\end{displaymath}

In addition, remembering that $O(\sqrt{S_{0}}) \leq c_{0} \leq O(S_{0})$ we can rewrite the
expression of Equation~\ref{eq.complexityDynamic1} as follows~:

\[
d \cdot O \left(
\begin{array}{c}
  c_{0} + dk \frac{\gamma_{1}}{c_{0}} + k \sqrt{S_{0}} + \frac{k^{2}}{\sqrt{S_{0}}} +  \\
  \sqrt{k c_{0}} \sqrt{ \frac{d}{c_{0}}\gamma_{1} + \frac{\sqrt{S_{0}}}{k} + \sqrt{S_{0}} + \frac{k}{\sqrt{S_{0}}}} + \sqrt{k d }
\end{array}
\right) =
\]
\[
d \cdot O \left(
\begin{array}{c}
  c_{0} + dk \frac{\gamma_{1}}{c_{0}} + k \sqrt{S_{0}} + \frac{k^{2}}{\sqrt{S_{0}}} + \\
  \sqrt{k c_{0}} \left(\sqrt{\gamma_{1}} + \sqrt[4]{S_{0}} \sqrt{\frac{\gamma_{1}}{k}} + \sqrt[4]{S_{0}} + \frac{\sqrt{k}}{\sqrt[4]{S_{0}}}\right) + \sqrt{kd}
\end{array}
\right)
\]

Using Lemma \ref{lemma.gamma1} we see that~:
\[
d \cdot O \left(
\begin{array}{c}
  c_{0} + dk \frac{\ln S_{0}}{c_{0}} + k \sqrt{S_{0}} + \frac{k^{2}}{\sqrt{S_{0}}} + \\
  \sqrt{c_{0} \ln S_{0} \sqrt{S_{0}}} +     \sqrt{c_{0} k \sqrt{S_{0}}} +   k \sqrt[4]{S_{0}} + \sqrt{k d}
\end{array}
\right) =
\]
\begin{equation}\label{eq.complexityDynamic2}
d \cdot O \left(
\begin{array}{c}
  c_{0} + k \sqrt{S_{0}} \ln S_{0} + \frac{k^{2}}{\sqrt{S_{0}}} + \\
  \sqrt{c_{0} \ln S_{0} \sqrt{S_{0}}} +     \sqrt{c_{0} k \sqrt{S_{0}}}
\end{array}
\right)
\end{equation}

Assuming that $k > O(\ln S_{0})$ and as $c_{0} = O(S_{0})$ we can now write~:
\begin{equation}\label{eq.complexityDynamic3}
O \left(
\begin{array}{c}
  \frac{S_{0}^{2.5}}{k} + S_{0}^{2} \ln S_{0} + S_{0}^{1.5} k \ln S_{0} + k^{2} \sqrt{S_{0}} \ln S_{0} + \\
  \frac{k^{3}}{\sqrt{S_{0}}}
  + \frac{S_{0}^{2.25}}{\sqrt{k}} + S_{0}^{1.75} \sqrt{k} + S_{0}^{0.75} k^{1.5}
\end{array}
\right)
\end{equation}

Using Equation \ref{eq.theorem.agents3} we can see that this translates to~:
\begin{equation}\label{eq.complexityDynamic4}
O \left( S_{0}^{1.5} k \ln S_{0} \right)
\end{equation}
\end{proof}
\end{theorem}

\section{Conclusions}

In this paper we have discussed the problem of cleaning or covering a connected region on the grid using a collaborate team of simple, finite-state-automata robotic agents. We have shown that when the regions are static, this can be done in $O(\frac{1}{k} \sqrt{n} \cdot c_{0} + c_{0} + k)$ time which equals $O(\frac{1}{k} n^{1.5} + n)$ time in the worst case, thus improving the previous results for this problem. In addition, we have shown that when the region is expanding in a constant rate (which is ``slow enough''), a team of $\Theta(\sqrt{n})$ robots can still be guaranteed to clean or cover it, in $O(n^{2} \ln n)$ time.

In addition, we have shown that teams of less than $\Omega(\sqrt{n})$ robots can \emph{never} cover a region that expands every $O(1)$ time steps, regardless of their sensing capabilities, communications and memory resources employed, or the algorithm used. As to regions that expand slower than every $O(1)$ time steps, we have shown the following impossibility results. First, a region of $n$ tiles that expands every $d$ time steps cannot be covered by a group of $k$ agents if $dk \leq O(\sqrt{n})$. Using Theorem \ref{theorem.thereisanswer} we can guarantee a coverage when $dk = \Omega(\frac{n^{1.5}}{\ln n})$, or even for $dk = \Omega(n))$ (when the region's perimeter $c_{0}$ equals $O(n)$).

Second, a spreading region cannot be covered when $d^{2}k$ is smaller than the order of the perimeter of the bounding rectangle of the region (which equals $O(n)$ in the worse case and $O(\sqrt{n})$ for shapes with small perimeters). Using Theorem \ref{theorem.thereisanswer} we can guarantee a coverage when $d^{2}k = \Omega(\frac{n^{2.5}}{\ln^{2} n})$, or for $d^{2}k = \Omega(n^{1.5})$ (when the region's perimeter $c_{0}$ equals $O(n)$).

We believe that these results can be easily applied to various other problems in which a team of agents are required to operate in an expanding grid domain. For example, this result can show that a team of $\Theta(\sqrt{n})$ cops can always catch a robber (or for that matter --- a group of robbers), moving (slowly) in an unbounded grid. Alternatively, robbers can be guaranteed to escape a team of less than $O(\sqrt{n})$ cops, if the area they are located in is unbounded.

\section{Vitae}

\begin{table}
\begin{tabular}{p{3cm}p{10cm}}
  \epsfig{figure=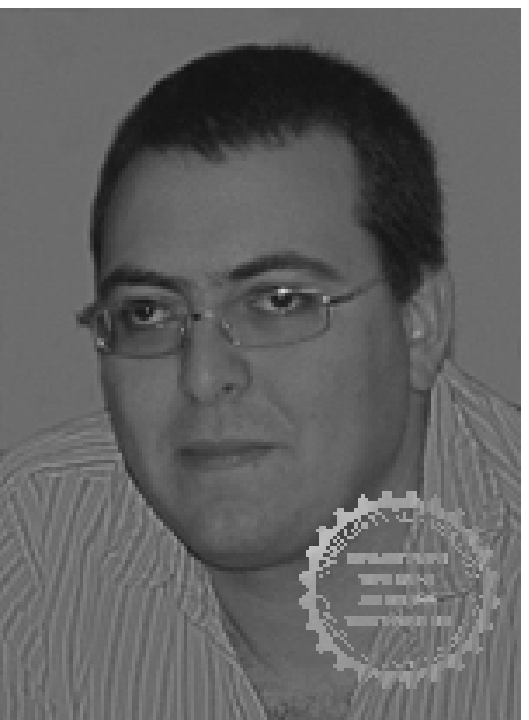,width=2.5cm}
  &
  Dr. Altshuler received his PhD in Computer Science from the Technion~---~Israel Institute of Technology (2009),
and is now a post-doc researcher at the Deutsche Telekom Research Lab in Ben Gurion University.  \\
  \epsfig{figure=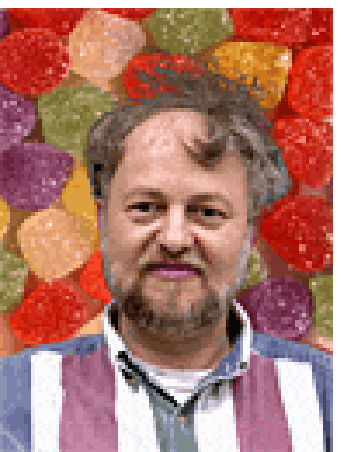,width=2.5cm}
  &
  Prof. Bruckstein is a professor at the Computer Science department at the Technion --- Israel Institute of Technology,
received his PhD in Electrical Engineering from Stanford University (1984).
\end{tabular}
\end{table}

\end{document}